\documentclass[12pt]{iopart}
\usepackage{graphicx}
\usepackage{iopams}
\eqnobysec

\begin{document}


\title[Low-energy expansion formula for asymptotically periodic potentials]
{Low-energy expansion formula 
for one-dimensional Fokker-Planck and Schr\"odinger equations
with asymptotically periodic potentials}

\author{Toru Miyazawa}

\address{Department of Physics, Gakushuin University, 
Tokyo 171-8588, Japan}
\ead{toru.miyazawa@gakushuin.ac.jp}
\begin{abstract}
We consider one-dimensional Fokker-Planck and Schr\"odinger equations with a potential which approaches a periodic function at spatial infinity. 
We extend the low-energy expansion method, which was introduced in previous papers, to be applicable to such asymptotically periodic cases. Using this method, we study the low-energy behavior of the Green function. 
\end{abstract}

\maketitle


\section{Introduction}
We consider the one-dimensional Fokker-Planck equation
\begin{equation}
\label{1-1.1}
-\frac{\rmd^2}{\rmd x^2}\phi(x)+2\frac{\rmd}{\rmd x}[f(x)\phi(x)]=k^2\phi(x)
\end{equation}
or the equivalent Schr\"odinger equation
\begin{equation}
\label{1-1.3}
-\frac{\rmd^2}{\rmd x^2}\psi(x)+V_{\rm S}(x)\psi(x)=k^2\psi(x).
\end{equation}
Equation~(1.1) describes the diffusion of particles in an external potential $V(x)$, from which the function $f(x)$ in (1.1) is defined by
\begin{equation}
\label{1-1.2}
f(x)=-\frac{1}{2}\frac{\rmd}{\rmd x}V(x).
\end{equation}
The Schr\"odinger potential $V_{\rm S}$ and the function $\psi$ of (1.2) are are related to $f$ and $\phi$ by
\begin{equation}
\label{1-1.4}
V_{\rm S}(x)=f^2(x)+f'(x)
\end{equation}
and
\begin{equation}
\phi(x)=\rme^{-V(x)/2}\psi(x).
\end{equation}
We shall always assume that ${\rm Im}\,k\geq 0$.
We define the Green function $G_{\rm S}(x,y;k)$ for the Schr\"odinger equation as the function satisfying
\begin{equation}
\label{1-1.5}
\left[\frac{\partial^2}{\partial x^2}-V_{\rm S}(x) +k^2 \right]G_{\rm S}(x,y;k)
=\delta(x-y)
\end{equation}
with the boundary conditions $G_{\rm S}(x,y;k)\to 0$ as $\vert x-y \vert \to \infty$ for ${\rm Im}\,k>0$. 
For ${\rm Im}\,k=0$, we define $G_{\rm S}(x,y;k)\equiv \lim_{\epsilon \downarrow 0}G_{\rm S}(x,y;k+\rm i\epsilon)$. 
Without loss of generality, we may suppose that $x \geq y$.

In a series of previous papers [1,2], we discussed a method for calculating the expansion of $G_{\rm S}(k)$ in powers of $k$.
In \cite{low}, we studied the cases in which the potential $V(x)$ either converges to a finite limit of diverges to infinity as $x \to \pm \infty$. 
In \cite{periodic}, we dealt with periodic potentials satisfying $V(x+L)=V(x)$. 
In this paper, we shall deal with asymptotically periodic potentials, i.e. potentials $V(x)$ that approach a periodic function as $x \to \pm \infty$. 
In solid state physics, impurities in a crystal are described by this type of potentials.  
The study of asymptotically periodic potentials is important for the application in physics, and there is a fair amount of literature on this subject [3-12].
However, there has not yet been a systematic analysis of the the low-energy behavior of the Green function up to high orders in $k$.
In this paper, we shall show that the method introduced in \cite{low} and \cite{periodic} can be extended to the asymptotically periodic case, enabling us to obtain the expansion of the Green function up to any order in $k$.

We assume that the potential $V(x)$ is a real-valued function which is piecewise continuously differentiable. (Note that $V(x)$ may have have jump discontinuities. See footnote~1 of \cite{periodic} and footnote~1 of \cite{low}.) 
We also assume that $V$ can be expressed as a sum of two functions: 
\begin{equation}
V(x)=V_{\rm p}(x)+ V_\Delta(x),
\end{equation}
where $V_{\rm p}$ is a periodic function satisfying
\numparts
\begin{equation}
V_{\rm p}(x+L)=V_{\rm p}(x),
\end{equation}
and $V_\Delta$ is a function such that
\begin{equation}
\lim_{x \to \pm \infty} V_\Delta(x)=0.
\end{equation}
\endnumparts
Corresponding to (1.7), we assume that the function $f$ (equation (1.3)) can be written as
\begin{equation}
f(x)=f_{\rm p}(x) + f_\Delta(x),
\end{equation}
where $f_{\rm p}(x+L)=f_{\rm p}(x)$ and $\lim_{x \to \pm \infty} f_\Delta(x)=0$. 

As we will see, the Green function can be expanded in terms of $k$ as 
\begin{eqnarray}
\fl
G_{\rm S}(x,y;k)=(\rmi k)^{-1}g_{-1}(x,y)+g_0(x,y)+\rmi k g_1(x,y)+ (\rmi k)^2 g_2(x,y)
\nonumber \\
+ \cdots + (\rmi k)^N g_N(x,y) + o(k^N)
\end{eqnarray}
if $V_\Delta \in L^1_N$, where $L^1_n$ denotes the set of functions $v(x)$ satisfying
\begin{equation}
\int_{-\infty}^\infty (1+ \vert x \vert^n)\vert v(x)\vert\, \rmd x < \infty.
\end{equation}
In this paper, we shall discuss the method for systematically calculating the coefficients $g_{-1}, g_0, g_1,\ldots$ of (1.10). 
Our method is based on the expansion formula for the reflection coefficient, which was derived in \cite{analysis}. 
This formula will be reviewed in section~3, after introducing some necessary notations in section~2. 
The calculation of $g_n$ is done in sections~4--9.

It is easy to extend this method to potentials $V(x)$ which have different asymptotic behaviors as $x \to -\infty$ and $x \to + \infty$. For example, we can easily deal with the cases where $V(x)$ approaches periodic functions with different periods as $x \to -\infty$ and $x \to +\infty$. This will be discussed in section~10.

If $V(x)$ is asymptotically periodic, the corresponding $V_{\rm S}(x)$ is also asymptotically periodic. That is to say, if $V$ has the form of (1.7) with (1.8), then $V_{\rm S}$ has the form
\begin{equation}
\fl
V_{\rm S}(x)=V_{\rm p}^{\rm S}(x)+ V_\Delta^{\rm S}(x), 
\qquad
V_{\rm p}^{\rm S}(x+L)=V_{\rm p}^{\rm S}(x), 
\qquad
\lim_{x \to \pm \infty}V_\Delta^{\rm S}(x)=0.
\end{equation}
But the converse is not necessarily true. 
Suppose that a Schr\"odinger potential $V_{\rm S}$ satisfying (1.12) is given. 
For simplicity, we assume that there are no bound states. 
Then the corresponding Fokker-Planck potential $V$ satisfies (1.7) and (1.8) only if the wave function $\psi(x)$ at the bottom of the lowest energy band remains finite for both $x \to + \infty$ and $x \to -\infty$. This is what is called the ^^ exceptional case' in the conventional terminology of scattering theory \cite{newton1}. 
In the ^^ generic case', the Fokker-Planck potential $V(x)$ corresponding to an asymptotically periodic $V_{\rm S}(x)$ is not asymptotically periodic but tends to $-\infty$ at either $x \to +\infty$ or $x \to -\infty$ (see example~2 in section~12). 
Our method is also applicable to $V_{\rm S}$ in the generic case, even though such $V_{\rm S}$ does not correspond to a Fokker-Planck potential satisfying (1.7) with (1.8).  
In the generic case, the expansion of $G_{\rm S}$ begins with the term of order $k^0$ (namely, $g_{-1}=0$ in (1.10)). 
In section~11, we will see how to calculate $g_0,g_1,g_2,\ldots$ for the generic case.


\section{Preliminaries}
Let the $2 \times 2$ matrix $U(x,x';k)$ be the solution of
\begin{equation}
\fl
\frac{\partial}{\partial x}
U(x,x';k)=
\left(
\begin{array}{cc}
-\rmi k & f(x) \\
f(x) & \rmi k \\
\end{array}
\right)
U(x,x';k),
\qquad
U(x',x';k)=
\left(
\begin{array}{cc}
1 & 0 \\
0 & 1 \\
\end{array}
\right).
\end{equation}
We write the elements of $U$ as
\begin{equation}
U(x,x';k)\equiv
\left(
\begin{array}{cc}
\alpha(x,x';k) & \beta(x,x';-k) \\
\beta(x,x';k) & \alpha(x,x';-k) \\
\end{array}
\right),
\end{equation}
and define the transmission coefficient $\tau$, the right reflection coefficient $R_r$ and the left reflection coefficient $R_l$ as
\refstepcounter{equation}
\begin{eqnarray}
\fl
\tau(x,x';k)\equiv \frac{1}{\alpha(x,x';k)},
\qquad
R_r(x,x';k)\equiv \frac{\beta(x,x';k)}{\alpha(x,x';k)},
\qquad
R_l(x,x';k)\equiv -\frac{\beta(x,x';-k)}{\alpha(x,x';k)}.
\nonumber \\
\fl
\end{eqnarray}
The generalized scattering coefficients $\bar \tau$, $\bar R_r$, $\bar R_l$ are defined with an additional variable $W$ as
\begin{eqnarray}
\fl
\bar \tau(x,x';W;k)\equiv 
\frac{\sqrt{1-\xi^2}\, \tau(x,x';k)}{1-\xi R_r(x,x';k)},
\qquad \bar R_r(x,x';W;k)\equiv 
\frac{R_r(x,x';k)-\xi}{1-\xi R_r(x,x';k)},
\nonumber \\
\fl
\bar R_l(x,x';W;k)\equiv 
R_l(x,x';k) + \frac{\xi\tau^2(x,x';k)}{1-\xi R_r(x,x';k)},
\end{eqnarray}
where
\begin{equation}
\xi(x,W)\equiv \tanh\frac{W-V(x)}{2}.
\end{equation}
(For an alternative definition, see (A.2) and (A.3) of appendix~A.)
We also define
\begin{equation}
S_r(x,k) \equiv \frac{R_r(x,-\infty;k)}{1+R_r(x,-\infty;k)}, 
\qquad
S_l(x,k) \equiv \frac{R_l(\infty,x;k)}{1+R_l(\infty,x;k)},
\end{equation}
\begin{equation}
S(x,k)\equiv S_r(x,k)+S_l(x,k).
\end{equation}
The Green function can be expressed in terms of this $S$ as \cite{expressions}
\begin{equation}
\fl
G_{\rm S}(x,y;k)=
\frac{1}{2\rmi k\sqrt{[1-S(x,k)][1-S(y,k)]}}
\exp\left[
\rmi k(x-y)-\rmi k\int_y^x S(z,k)\,\rmd z
\right].
\end{equation}

We use the notation
\begin{equation}
\label{1-2.15}
\fl
[\sigma_1,\sigma_2,\ldots,\sigma_n]_a^b\equiv \int \cdots \int_{a \leq z_1\leq z_2 \leq \cdots \leq z_n\leq b}
\rmd z_1 \rmd z_2\cdots \rmd z_n
\exp\Biggl[
\sum_{j=0}^n \sigma_j V(z_j)
\Biggr],
\end{equation}
for $n=1,2,3,\ldots$ and $-\infty\leq a\leq b\leq \infty$, where each $\sigma_j$ is either $+1$ or $-1$.
For simplicity, we write $[\hbox{$+$}]_a^b$, $[\hbox{$+$}\hbox{$-$}]_a^b$, etc in place of $[+1]_a^b$, $[+1,-1]_a^b$, etc.  
In this paper, we also deal with integrals of the form of (2.10) with $V$ replaced by $V_{\rm p}$. We will denote such integrals with a left subscript ^^ p' as 
\begin{equation}
\fl
{}_{\rm p}[\sigma_1,\sigma_2,\ldots,\sigma_n]_a^b\equiv \int \cdots \int_{a \leq z_1\leq z_2 \leq \cdots \leq z_n\leq b}
\rmd z_1 \rmd z_2\cdots \rmd z_n
\exp\Biggl[
\sum_{j=0}^n \sigma_j V_{\rm p}(z_j)
\Biggr].
\end{equation}
As in \cite{periodic}, we use the symbols
\begin{eqnarray}
\fl
P \equiv {}_{\rm p}[\hbox{$+$}]_{x-L}^x,
\qquad 
M \equiv {}_{\rm p}[\hbox{$-$}]_{x-L}^x,
\qquad
L_0 \equiv \sqrt{P M},
\qquad
V_0 \equiv \frac{1}{2} \log \frac{P}{M},
\nonumber \\
Q \equiv {}_{\rm p}[\hbox{$-$$+$$-$$+$}]_{x-L}^x+{}_{\rm p}[\hbox{$+$$-$$+$$-$}]_{x-L}^x.
\end{eqnarray}
All the quantities defined in (2.12) are independent of $x$.

\section{Formula for the expansion of $\boldsymbol{\bar R_r(x,-\infty;W;k)}$}
In this section, we summarize the necessary results from \cite{analysis}.
We define the operators $\mathcal{A}$ and $\mathcal{B}$, which act on functions of $x$ and $W$, as
\begin{eqnarray}
\label{1-2.21}
\mathcal{A}h(x,W) \equiv \frac{\partial}{\partial x}h(x,W),
\\
\label{1-2.22}
\mathcal{B}h(x,W) 
\equiv \frac{\partial}{\partial W}
\left\{ \sinh[W-V(x)]h(x,W)\right\}.
\end{eqnarray}
The generalized reflection coefficient $\bar R_r(x,-\infty;W;k)$ satisfies the differential equation
\begin{equation}
(\mathcal{A}-2\rmi k\mathcal{B})
\left[
\bar R_r(x,-\infty;W;k) + \xi(x,W)
\right]
=
\left[
1- \xi^2(x,W)
\right]
f(x).
\end{equation}
From (3.3) we have
\begin{equation}
\label{1-3.7}
\bar R_r(x,-\infty;W;k)
=-\xi+(\mathcal{A}-2\rmi k\mathcal{B})^{-1}(1-\xi^2)f,
\end{equation}
where $(\mathcal{A}-2\rmi k\mathcal{B})^{-1}$ is the inverse of the operator $\mathcal{A}-2\rmi k\mathcal{B}$.
This inverse is given by
\begin{equation}
\label{1-3.6}
\fl
(\mathcal{A}-2 \rmi k\mathcal{B})^{-1} g(x,W)
=\int_{-\infty}^x \rmd z\, \frac{\bar \tau^2(x,z;W;k)}{1-\bar R_l^2(x,z;W;k)}
g(z,\bar \omega(x,z;W;k)),
\end{equation}
where
\begin{equation}
\label{1-3.5}
\bar \omega(x,z;W;k) \equiv V(z)+\log \frac{1+\bar R_l(x,z;W;k)}{1-\bar R_l(x,z;W;k)}.
\end{equation}
We can formally expand $(\mathcal{A}-2 \rmi k\mathcal{B})^{-1}$ in powers of $k$ as
\begin{eqnarray}
\fl
(\mathcal{A}-2 \rmi k\mathcal{B})^{-1}&=
\left[1+ \rmi k (2\mathcal{A}^{-1} \mathcal{B}) + (\rmi k)^2 (2\mathcal{A}^{-1} \mathcal{B})^2 + \cdots + (\rmi k)^N (2\mathcal{A}^{-1} \mathcal{B})^N\right]\mathcal{A}^{-1}
\nonumber \\
&\qquad +(\rmi k)^{N+1}(\mathcal{A}-2 \rmi k\mathcal{B})^{-1}\mathcal{A} (2\mathcal{A}^{-1} \mathcal{B})^{N+1} \mathcal{A}^{-1}.
\end{eqnarray}
Substituting (3.7) into (3.4) yields the expansion
\begin{equation}
\label{1-3.11}
\bar R_r=\bar r_0 + \rmi k \bar r_1 + (\rmi k)^2 \bar r_2 + \cdots + (\rmi k)^N \bar r_N+\bar \rho_N
\end{equation}
with
\refstepcounter{equation}
\label{1-3.12}
\addtocounter{equation}{-1}
\numparts
\begin{equation}
\bar r_0=\mathcal{A}^{-1}(1-\xi^2)f - \xi,
\end{equation}
\begin{equation}
\bar r_n=(2\mathcal{A}^{-1} \mathcal{B})^n(\bar r_0 +\xi) \qquad (n \geq 1),
\end{equation}
\endnumparts
\begin{equation}
\fl
\bar \rho_N 
=(\rmi k)^{N+1}(\mathcal{A}-2 \rmi k\mathcal{B})^{-1}\mathcal{A}\,\bar r_{N+1}
=2(\rmi k)^{N+1}(\mathcal{A}-2 \rmi k\mathcal{B})^{-1}\mathcal{B}\,\bar r_N.
\end{equation}
Equation (3.8) makes sense if and only if the right-hand side of (3.9{\it b}) makes sense for all $n \leq N$. (The remainder term $\bar \rho_N$ automatically makes sense if all $\bar r_n$ make  sense.)

\section{Inverse of the operator $\mathcal{A}$}
In the formal expression (3.7), the symbol $\mathcal{A}^{-1}$ denotes the inverse of $\mathcal{A}$. 
However, ^^ inverse of $\mathcal{A}$' does not have a meaning unless we specify the domain of $\mathcal{A}$ . 
Specifying the domain of $\mathcal{A}$ amounts to specifying the boundary condition for $\bar R_r(x,-\infty;W;k)$ at $x \to -\infty$. 
Since the operator $\mathcal{A}$ in (3.3) acts on the function $\bar R_r(x,-\infty;W;k) + \xi(x,W)$, the domain of $\mathcal{A}$ must be chosen so that $\bar R_r + \xi$ belongs to that domain.

We define the operators 
$\mathcal{A}_0^{-1}$ and $\mathcal{A}_{\rm p}^{-1}$ by
\begin{eqnarray}
\fl
\mathcal{A}_0^{-1}g(x,W) \equiv \int_{-\infty}^x g(z,W)\,\rmd z,
\\
\fl
\mathcal{A}_{\rm p}^{-1}g(x,W)
\equiv 
\frac{1}{L_0 \sinh(W-V_0)}
\nonumber \\
\times
\int_{x-L}^x \rmd z \int_z^x\rmd z'
\Bigl\{\sinh[V_{\rm p}(z')-W] g(z,W)
 -\sinh[V_{\rm p}(z')-V_0]g(z,V_0) \Bigr\}.
 \nonumber \\
 \fl
\end{eqnarray}
Let $D_\Delta$ denote the set of two-variable functions $h(x,W)$ which are piecewise continuously differentiable with respect to $x$, analytic with respect to $W$ on the real axis and which satisfy
\begin{equation}
\lim_{x \to -\infty} h(x,W)=0.
\end{equation}
And let $D_{\rm p}$ denote the set of functions $h(x,W)$ which are piecewise continuously differentiable with respect to $x$, analytic with respect to $W$ on the real axis and which satisfy the conditions
\begin{equation}
\fl
h(x+L,W)=h(x,W)
\quad
\hbox{and}
\quad
\int_{x-L}^x \mathcal{B}_{\rm p} h(z,W) \,\rmd z=0
\quad
\hbox{for any $x$,}
\end{equation}
where we have defined, corresponding to (3.2), 
\begin{equation}
\mathcal{B}_{\rm p}h(x,W) 
\equiv \frac{\partial}{\partial W}
\left\{ \sinh[W-V_{\rm p}(x)]h(x,W)\right\}.
\end{equation}
It is easy to see that
\numparts
\begin{equation}
\mathcal{A}_0^{-1} \mathcal{A} h=h
\quad \hbox{if} \quad h \in D_\Delta.
\end{equation}
(We are allowing $\mathcal{A} h$ to include delta functions.) 
And it was shown in \cite{periodic} that 
\begin{equation}
\mathcal{A}_{\rm p}^{-1} \mathcal{A} h=h
\quad \hbox{if} \quad h \in D_{\rm p}.
\end{equation} 
\endnumparts
(In \cite{periodic}, $\mathcal{A}_{\rm p}^{-1}$, $\mathcal{B}_{\rm p}$ etc are written without the subscript ${\rm p}$.) In other words, if the domain of $\mathcal{A}$ is restricted to $D_\Delta$ or $D_{\rm p}$, the inverse of $\mathcal{A}$ is given by $\mathcal{A}_0^{-1}$ or $\mathcal{A}_{\rm p}^{-1}$,  respectively.

If $\bar R_r(x,-\infty;W;k)+\xi(x,W)$ belongs to $D_\Delta$, the domain of the operator $\mathcal{A}$ in (3.3) can be taken to be $D_\Delta$. Then the inverse of $\mathcal{A}$ is given by (4.1), and the expansion of $\bar R_r$ is obtained by letting $\mathcal{A}^{-1}=\mathcal{A}_0^{-1}$ in (3.9). This is the case for the non-periodic potentials discussed in \cite{low}. 
On the other hand, if the potential is periodic, then $\bar R_r + \xi$ belongs to $D_{\rm p}$ (see \cite{periodic} for details), and so $\mathcal{A}_{\rm p}^{-1}$ should be used in place of $\mathcal{A}^{-1}$ in (3.9). 
In this paper, we are assuming that the potential has the form of (1.7). 
In order to use the expansion formula shown in the previous section, we must find an appropriate domain of $\mathcal{A}$ for such asymptotically periodic potentials.

Let $\bar R_r^{\rm p}$ denote $\bar R_r$ with $V$ replaced by $V_{\rm p}$. (That is to say, $\bar R_r^{\rm p}$ is defined in the same way as the definition of $\bar R_r$ in (2.1)--(2.5), with $f$ in (2.1) replaced by $f_{\rm p}$.) We define $\bar R_r^\Delta \equiv \bar R_r -\bar R_r^{\rm p}$, and
\begin{equation}
\xi_{\rm p}(x,W)\equiv \tanh\frac{W-V_{\rm p}(x)}{2},
\qquad
\xi_\Delta(x,W) \equiv \xi(x,W) - \xi_{\rm p}(x,W).
\end{equation}
Then, we can write
\begin{eqnarray}
\bar R_r + \xi = \bar R_r^{\rm p} + \xi_{\rm p} + \bar R_r^\Delta + \xi_\Delta.
\end{eqnarray}
Both $\bar R_r^\Delta(x,-\infty;W;k)$ and $\xi_\Delta(x,W)$ vanish as $x \to -\infty$, and hence it can be seen that $\bar R_r^\Delta + \xi_\Delta \in D_\Delta$. 
As for the periodic part, it was shown in \cite{periodic} that $\bar R_r^{\rm p}+ \xi_{\rm p}\in D_{\rm p}$. Therefore, $\bar R_r+\xi \in D_{\rm p} + D_\Delta$
(where $h \in D_{\rm p} + D_\Delta$ means that $h= h_{\rm p} + h_\Delta$ with $h_{\rm p} \in D_{\rm p}$ and $h_\Delta \in D_\Delta$).
So we know that we should take $D_{\rm p}+ D_\Delta$ as the domain of $\mathcal{A}$.

From now on, we assume that the operator $\mathcal{A}$ is defined with the domain $D_{\rm p}+ D_\Delta$. 
To calculate (3.9), we need the inverse of $\mathcal{A}$.
If $h \in D_{\rm p} + D_\Delta$ and $g = \mathcal{A} h$, we can write
\begin{equation}
g=g_{\rm p}+ g_\Delta,
\end{equation}
where $g_{\rm p}=\mathcal{A}h_{\rm p}$ and $g_\Delta=\mathcal{A}h_\Delta$ with $h_{\rm p} \in D_{\rm p}$ and $h_\Delta \in D_\Delta$. 
Any function $g$ belonging to the range of $\mathcal{A}$ can be uniquely decomposed into two parts as (4.9), where $g_{\rm p}$ is a function satisfying
\begin{equation}
g_{\rm p}(x+L,W)=g_{\rm p}(x,W),
\quad
\int_{x-L}^x\! g_{\rm p}(z,W)\,\rmd z=0
\quad \hbox{for any $x$},
\end{equation}
and $g_\Delta$ is a function such that
\begin{equation}
\int_{-\infty}^x g_\Delta(z,W)\,\rmd z
\quad
\hbox{exists and is finite.}
\end{equation}
It is obvious that $g_{\rm p}$ satisfies (4.10) if $g_{\rm p}=\mathcal{A}h_{\rm p}$ with $h_{\rm p} \in D_{\rm p}$. Conversely, if $g_{\rm p}$ satisfies (4.10), it can be shown that $\mathcal{A}_{\rm p}^{-1} g_{\rm p} \in D_{\rm p}$ and $\mathcal{A}\mathcal{A}_{\rm p}^{-1} g_{\rm p}=g_{\rm p}$ (see \cite{periodic}). 
From $\mathcal{A}^{-1}g=h_{\rm p}+h_\Delta$ and (4.6), we have
\begin{equation}
\mathcal{A}^{-1} g=\mathcal{A}_{\rm p}^{-1} g_{\rm p} + \mathcal{A}_0^{-1} g_\Delta.
\end{equation}
Thus, when the potential is asymptotically periodic, the expansion of $\bar R_r$ is given by (3.8) and (3.9) with $\mathcal{A}^{-1}$ acting as (4.12). To calculate $\mathcal{A}^{-1}g$, we first express $g$ as the sum of the periodic part $g_{\rm p}$ and the non-periodic part $g_\Delta$, satisfying  (4.10) and (4.11), respectively. Then, the right-hand side of (4.12) is calculated with the operators defined by (4.1) and (4.2).

\section{Expressions for $\boldsymbol{\bar r_n}$}
Now let us calculate the coefficients $\bar r_n$ given by (3.9).
To calculate $\bar r_0$, it is necessary to decompose $(1-\xi^2) f$ into periodic and non-periodic parts.
Note that
\begin{equation}
[1-\xi^2(x,W)]f(x)=
\frac{-1}{2 \cosh^2 \frac{W-V(x)}{2}} 
\frac{\rmd }{\rmd x}V(x) 
=\frac{\partial}{\partial x}\xi(x,W).
\end{equation}
Using (4.7), we write (5.1) as
\begin{equation}
[1-\xi^2(x,W)]f(x)=\frac{\partial}{\partial x}\xi_{\rm p}(x,W)+\frac{\partial}{\partial x}\xi_\Delta(x,W).
\end{equation}
The two terms on the right-hand side of (5.2) correspond, respectively, to $g_{\rm p}$ and $g_\Delta$ of (4.9). They satisfy conditions (4.10) and (4.11). 
So, according to (4.12),
\begin{equation}
\mathcal{A}^{-1}(1-\xi^2) f
=\mathcal{A}_{\rm p}^{-1}
\frac{\partial}{\partial x}\xi_{\rm p}(x,W)
+\mathcal{A}_0^{-1}
\frac{\partial}{\partial x}\xi_\Delta(x,W).
\end{equation}
The first term on the right-hand side has already been calculated in \cite{periodic}. The result is
\begin{equation}
\mathcal{A}_{\rm p}^{-1}
\frac{\partial}{\partial x}\xi_{\rm p}(x,W)
=\xi_{\rm p}(x,W)-\tanh \frac{W-V_0}{2}
\end{equation}
(see equation~(8.1) of \cite{periodic}). 
It is obvious that the second term of (5.3) is
\begin{equation}
\fl
\mathcal{A}_0^{-1}
\frac{\partial}{\partial x}\xi_\Delta(x,W)
=\int_{-\infty}^x \rmd z
\,\frac{\partial}{\partial z}\xi_\Delta(z,W)
=\xi_\Delta(x,W)
=\xi(x,W)-\xi_{\rm p}(x,W).
\end{equation}
(Note that 
$\mathcal{A}_0^{-1}\frac{\partial}{\partial x}\xi_\Delta 
= \xi_\Delta$ 
but
$\mathcal{A}_{\rm p}^{-1}\frac{\partial}{\partial x}\xi_{\rm p}\neq \xi_{\rm p}$
 since $\xi_\Delta \in D_\Delta$ but $\xi_{\rm p} \notin D_{\rm p}$.)
Therefore,
\begin{equation}
\mathcal{A}^{-1}(1-\xi^2) f
=-\tanh \frac{W-V_0}{2} + \xi(x,W),
\end{equation}
and (3.9{\it a}) gives
\begin{equation}
\bar r_0=-\tanh \frac{W-V_0}{2}.
\end{equation}

Let us proceed to the calculation of $\bar r_1= 2 \mathcal{A}^{-1}\mathcal{B} (\bar r_0 + \xi)$. It can be shown that
\begin{equation}
2\mathcal{B}(\bar r_0 + \xi)
=\frac{1}{\cosh^2 \frac{W-V_0}{2}} \sinh[V_0-V(x)]
\end{equation}
(see equation~(8.3) of \cite{periodic}).
We decompose the right-hand side as
\begin{equation}
2\mathcal{B}(\bar r_0 + \xi)
=\bar q_0^{\rm p}+\bar q_0^\Delta,
\end{equation}
where
\begin{equation}
\fl
\bar q_0^{\rm p} \equiv \frac{1}{\cosh^2 \frac{W-V_0}{2}} 
\sinh[V_0-V_{\rm p}(x)],
\qquad
\bar q_0^\Delta \equiv \frac{1}{\cosh^2 \frac{W-V_0}{2}} 
\sinh[V_0-V(x)] -\bar q_0^{\rm p}.
\end{equation}
It is easy to check that $\bar q_0^{\rm p}$ satisfies conditions (4.10). In order that $\bar q_0^\Delta$ satisfy (4.11), it is necessary that $V(x)$ approach $V_{\rm p}(x)$ sufficiently rapidly as $x \to -\infty$ (see the next section).
Assuming that this condition is satisfied, we have, from (5.9) and (4.12),
\begin{equation}
2\mathcal{A}^{-1}\mathcal{B}(\bar r_0 + \xi)=\mathcal{A}_{\rm p}^{-1}\bar q_0^{\rm p} + \mathcal{A}_0^{-1}\bar q_0^\Delta.
\end{equation}
So we obtain $\bar r_1$ as a sum of two terms
\begin{equation}
\bar r_1=\bar r_1^{\rm p}+ \bar r_1^\Delta,
\end{equation}
where 
\numparts
\begin{eqnarray}
\bar r_1^{\rm p}
\equiv
\mathcal{A}_{\rm p}^{-1}\frac{1}{\cosh^2 \frac{W-V_0}{2}} 
\sinh[V_0-V_{\rm p}(x)],
\\
\bar r_1^\Delta\equiv
\mathcal{A}_0^{-1}\frac{1}{\cosh^2 \frac{W-V_0}{2}} 
 \biggl\{\sinh[V_0-V(x)]-\sinh[V_0-V_{\rm p}(x)]\biggr\}.
\end{eqnarray}
\endnumparts
The right-hand side of (5.13{\it a}) can be calculated using (4.2). 
Details for the calculation is given in \cite{periodic}. 
As a result, we have
\begin{equation}
\bar r_1^{\rm p}
=\frac{1}{4 L_0 \cosh^2 \frac{W-V_0}{2}}
\Bigl(
{}_{\rm p}[\hbox{$+$$-$}]_{x-L}^x-{}_{\rm p}[\hbox{$-$$+$}]_{x-L}^x
\Bigr).
\end{equation}
(See equation~(8.7) of \cite{periodic}. Note that $[\sigma_1,\ldots,\sigma_n]_a^b$ in \cite{periodic} is ${}_{\rm p}[\sigma_1,\ldots,\sigma_n]_a^b$ in this paper.)
Equation (5.13{\it b}) can be written as
\begin{eqnarray}
\bar r_1^\Delta
=\frac{1}{2 \cosh^2 \frac{W-V_0}{2}}
\int_{-\infty}^x
\left[
\rme^{V_0} \Delta^-(z) - \rme^{-V_0} \Delta^+(z)
\right] \rmd z,
\end{eqnarray}
where we have defined 
\begin{equation}
\Delta^{\pm}(x) \equiv \rme^{\pm V(x)}-\rme^{\pm V_{\rm p}(x)}.
\end{equation}
Obviously $\bar r_1^\Delta$ vanishes as $x \to -\infty$ (provided that the integral on the right-hand side of (5.15) is convergent), while $\bar r_1^{\rm p}$ is a periodic function of $x$. 
The first-order coefficient $\bar r_1$ is thus obtained as the sum of the periodic part (5.14) and the non-periodic part (5.15).

To calculate $\bar r_n$ for larger $n$, we can use the recursion relation $\bar r_n =2 \mathcal{A}^{-1}\mathcal{B} \bar r_{n-1}$, which follows from (3.9{\it b}). We define
\begin{equation}
\bar q_n \equiv 2 \mathcal{B} \bar r_n  \quad (n\geq 1),
\end{equation}
so that
\begin{equation}
\bar r_n=\mathcal{A}^{-1} \bar q_{n-1}.
\end{equation}
We assume that $\bar r_n$ and $\bar q_n$ can be written as the sum of periodic and non-periodic parts,
\begin{equation}
\bar r_n=\bar r_n^{\rm p}+\bar r_n^\Delta,
\qquad
\bar q_n= \bar q_n^{\rm p} + \bar q_n^\Delta,
\end{equation}
where
\numparts
\begin{eqnarray}
\bar r_n^{\rm p}(x+L;W)= \bar r_n^{\rm p}(x,W), 
\qquad
\lim_{x \to -\infty} \bar r_n^\Delta(x,W)=0,
\\
\bar q_n^{\rm p}(x+L;W)= \bar q_n^{\rm p}(x,W), 
\qquad
\lim_{x \to -\infty} \bar q_n^\Delta(x,W)=0.
\end{eqnarray}
\endnumparts
This assumption will be justified by the result.
We split $\mathcal{B}$ into two parts as
\begin{equation}
\mathcal{B}=\mathcal{B}_{\rm p} +\mathcal{B}_\Delta,
\end{equation}
where $\mathcal{B}_{\rm p}$ is defined by (4.5), and 
$\mathcal{B}_\Delta (\equiv \mathcal{B}-\mathcal{B}_{\rm p})$ can be expressed in terms of $\Delta^\pm$ (equation (5.16)) as
\begin{eqnarray}
\mathcal{B}_\Delta 
=\frac{\rme^W}{2}
 \Delta^-(x) \left(1+\frac{\partial}{\partial W}\right)
+\frac{\rme^{-W}}{2} \Delta^+(x) \left(1-\frac{\partial}{\partial W}\right).
\end{eqnarray}
From (5.17), (5.19) and (5.21), it follows that
\begin{equation}
\bar q_n^{\rm p}=2\mathcal{B}_{\rm p}\, \bar r_n^{\rm p},
\qquad
\bar q_n^\Delta=2 \mathcal{B}\,\bar r_n^\Delta
+ 2 \mathcal{B}_\Delta \,\bar r_n^{\rm p}.
\end{equation}
It can be shown \cite{periodic} that $\bar q_n^{\rm p}$ satisfies conditions (4.10), and so $\mathcal{A}_{\rm p}^{-1} \bar q_n^{\rm p}$ makes sense. Assuming that $\mathcal{A}_0^{-1} \bar q_n^\Delta$ also makes sense, we have $\mathcal{A}^{-1}\bar q_n=\mathcal{A}_{\rm p}^{-1}\bar q_n^{\rm p} + \mathcal{A}_0^{-1}\bar q_n^\Delta$. Therefore,
\begin{equation}
\fl
\bar r_n^{\rm p}
=\mathcal{A}_{\rm p}^{-1}\bar q_{n-1}^{\rm p}
=2 \mathcal{A}_{\rm p}^{-1}\mathcal{B}_{\rm p}\, \bar r_{n-1}^{\rm p},
\qquad
\bar r_n^\Delta
=\mathcal{A}_0^{-1}\bar q_{n-1}^\Delta
=2 \mathcal{A}_0^{-1}\mathcal{B}\,\bar r_{n-1}^\Delta
+ 2 \mathcal{A}_0^{-1}\mathcal{B}_\Delta\, \bar r_{n-1}^{\rm p}.
\end{equation}
By iterating (5.24), we obtain
\begin{eqnarray}
\bar r_n^{\rm p}=(2 \mathcal{A}_{\rm p}^{-1} \mathcal{B}_{\rm p})^{n-1} \bar r_1^{\rm p},
\\
\bar r_n^\Delta=(2 \mathcal{A}_0^{-1} \mathcal{B})^{n-1} \bar r_1^\Delta
+\sum_{j=0}^{n-2}(2 \mathcal{A}_0^{-1} \mathcal{B})^j\,
2 \mathcal{A}_0^{-1}\mathcal{B}_\Delta \,\bar r_{n-j-1}^{\rm p}.
\end{eqnarray}
In this way, $\bar r_n^{\rm p}$ and $\bar r_n^\Delta$ can be calculated from $\bar r_1^{\rm p}$ and $\bar r_1^\Delta$ (equations~(5.14) and (5.15)). 
It is obvious from (5.25) and (5.26) that equations (5.20a) are satisfied as long as the right-hand side of (5.26) makes sense.
If $V_\Delta=0$, then $\bar r_n=\bar r_n^{\rm p}$. 
The $\bar r_n^{\rm p}$ are identical with the $\bar r_n$ studied in \cite{periodic} for periodic potentials. 

Let us explicitly write out the expression for $\bar r_2=\bar r_2^{\rm p}+ \bar r_2^\Delta$. The calculation of $\bar r_2^{\rm p}$ was done in \cite{periodic}. The result is given by equation (8.18) of \cite{periodic} as
\begin{eqnarray}
\fl
\bar r_2^{\rm p}
=\frac{1}{4 L_0 \cosh^3 \frac{W-V_0}{2}}
\Biggl\{
\rme^{-(W+V_0)/2} {}_{\rm p}[\hbox{$+$$-$$+$}]_{x-L}^x
-\rme^{(W+V_0)/2} {}_{\rm p}[\hbox{$-$$+$$-$}]_{x-L}^x
\nonumber \\
\qquad
+ 
\frac{1}{4 L_0} \left(L_0^4 + 4 Q \right) \sinh \frac{W-V_0}{2}
\Biggr\},
\end{eqnarray}
with $Q$ defined by (2.12). 
For $n=2$, the right-hand side of (5.26) consists of the two terms $2 \mathcal{A}_0^{-1}\mathcal{B}\bar r_1^\Delta$ and $2 \mathcal{A}_0^{-1}\mathcal{B}_\Delta \bar r_1^{\rm p}$. Applying (5.22) and (3.2) to (5.14) and (5.15) respectively, we can calculate
\numparts
\begin{eqnarray}
\fl
\mathcal{B}_\Delta \bar r_1^{\rm p}
= \frac{\rme^{(W+V_0)/2}\Delta^-(x) +\rme^{-(W+V_0)/2}\Delta^+(x)}{8 L_0 \cosh^3 \frac{W-V_0}{2}}
\Bigl(
{}_{\rm p}[\hbox{$+$$-$}]_{x-L}^x-{}_{\rm p}[\hbox{$-$$+$}]_{x-L}^x
\Bigr),
\\
\fl
\mathcal{B}\bar r_1^\Delta
=\frac{\rme^{(W+V_0)/2}\,\rme^{-V(x)} + \rme^{-(W+V_0)/2}\,\rme^{V(x)}}
{4\cosh^3 \frac{W-V_0}{2}}
\int_{-\infty}^x 
\left[
\rme^{V_0} \Delta^-(z)- \rme^{-V_0} \Delta^+(z)
\right]
\rmd z.
\end{eqnarray}
\endnumparts
Hence, the non-periodic part $\bar r_2^\Delta$ is obtained as
\begin{eqnarray}
\fl
\bar r_2^\Delta=\frac{1}{4 L_0 \cosh^3 \frac{W-V_0}{2}}
\nonumber \\
\times
\int_{-\infty}^x 
\Biggl\{
\left[
\rme^{(W+V_0)/2}\Delta^-(z) +\rme^{-(W+V_0)/2}\Delta^+(z)
\right]
\Bigl(
{}_{\rm p}[\hbox{$+$$-$}]_{z-L}^z-{}_{\rm p}[\hbox{$-$$+$}]_{z-L}^z
\Bigr)
\nonumber \\
\quad
+
2L_0
\Bigl(
\rme^{(W+V_0)/2}[\hbox{$-$}]_z^x +\rme^{-(W+V_0)/2}[\hbox{$+$}]_z^x
\Bigr)
\left[
\rme^{V_0} \Delta^-(z) - \rme^{-V_0} \Delta^+(z)
\right]
\Biggr\}\, \rmd z,
\nonumber \\
\end{eqnarray}
where we have used
$\int_{-\infty}^x \rmd z \int_{-\infty}^z \rmd z'=\int_{-\infty}^x \rmd z' \int_{z'}^x \rmd z$.

\section{Condition for the finiteness of $\boldsymbol{\bar r_n}$}

Equation (3.8) is meaningless unless the coefficients $\bar r_n=\bar r_n^{\rm p}+\bar r_n^\Delta$ are finite for all $n \leq N$.  The periodic part $\bar r_n^{\rm p}$, which is given by (5.25), is finite for any $n$ (see \cite{periodic}). 
However, the terms on the right-hand side of (5.26) are not necessarily finite. For these terms to be finite, it is necessary that $V_\Delta(x)$ vanish sufficiently fast as $x\to -\infty$. 

Since $\Delta^\pm (z) \sim \pm \rme^{V_{\rm p}(z)} V_\Delta (z)$ as $z \to -\infty$,
and since $V_{\rm p}(z)$ is a bounded periodic function, there exists a constant $C$ such that
\begin{equation}
\left\vert
\Delta^\pm (z)
\right\vert
\leq C \left\vert V_\Delta(z)\right\vert
\end{equation}
for $-\infty<z<x$ with fixed $x$. Using this in (5.15), we find
\begin{equation}
\left\vert \bar r_1^\Delta \right\vert
\leq
C \int_{-\infty}^x \left\vert V_\Delta(z) \right\vert \rmd z.
\end{equation}
(We will use the symbol $C$ to denote a finite constant which is not necessarily the same at each occurrence. We regard $x$ and $W$ as fixed.) 
We can see from (5.28{\it b}) that
\begin{equation}
\fl
\left\vert 2 \mathcal{A}_0^{-1} \mathcal{B}\, \bar r_1^\Delta\right\vert 
\leq
C
\int_{-\infty}^x \rmd z
\int_{-\infty}^z \rmd z'
\left\vert
\rme^{V_0} \Delta^-(z') - \rme^{-V_0} \Delta^+(z')
\right\vert,
\end{equation}
since $\rme^{\pm V(z)}$ are bounded for $-\infty<z<x$.
Using (6.1) in (6.3) gives
\begin{eqnarray}
\fl
\left\vert 2 \mathcal{A}_0^{-1} \mathcal{B}\, \bar r_1^\Delta\right\vert 
\leq
C
\int_{-\infty}^x \rmd z
\int_{-\infty}^z \rmd z'
\left\vert V_\Delta(z')\right\vert
=
C
\int_{-\infty}^x \rmd z'
\int_{z'}^x \rmd z
\left\vert V_\Delta(z')\right\vert
\nonumber \\
=C
\int_{-\infty}^x
(x-z)
\left\vert
V_\Delta(z)
\right\vert \rmd z.
\end{eqnarray}
More generally, it is easy to see that
\begin{equation}
\fl
\left\vert (2 \mathcal{A}_0^{-1} \mathcal{B})^{n-1}\, \bar r_1^\Delta \right\vert
\leq
C \left\vert (\mathcal{A}_0^{-1})^{n-1}\, \bar r_1^\Delta \right\vert
\leq
C \,(\mathcal{A}_0^{-1})^{n-1} 
\int_{-\infty}^x \left\vert V_\Delta(z) \right\vert \rmd z.
\end{equation}
Hence we can derive, in the same way as (6.4),
\begin{equation}
\left\vert (2 \mathcal{A}_0^{-1} \mathcal{B})^{n-1}\, \bar r_1^\Delta \right\vert
\leq
C
\int_{-\infty}^x
(x-z)^{n-1}
\left\vert
V_\Delta(z)
\right\vert \rmd z.
\end{equation}
This gives an upper bound for the first term on the right-hand side of (5.26). 
Similar inequalities hold for the remaining terms of (5.26).
Using (5.22) and (6.1), and also using the fact that $\vert \bar r_{n-j-1}^{\rm p}\vert \leq C$ and $\vert \frac{\partial}{\partial W}\bar r_{n-j-1}^{\rm p}\vert \leq C$, we have
\begin{equation}
\left\vert
2 \mathcal{A}_0^{-1}\mathcal{B}_\Delta \bar r_{n-j-1}^{\rm p}
\right\vert
\leq
C
\int_{-\infty}^x
\left\vert
V_\Delta(z)
\right\vert \rmd z,
\end{equation}
and, just like (6.6), 
\begin{equation}
\left\vert
(2 \mathcal{A}_0^{-1} \mathcal{B})^j \,2 \mathcal{A}_0^{-1}\mathcal{B}_\Delta \bar r_{n-j-1}^{\rm p}
\right\vert
\leq
C
\int_{-\infty}^x
(x-z)^j
\left\vert
V_\Delta(z)
\right\vert \rmd z.
\end{equation}

Let $F^{(-)}_n$ denote the set of functions $v(x)$ which satisfy the condition
\begin{equation}
\int_{-\infty}^a (1+ \vert x \vert^n)\vert v(x)\vert\, \rmd x < \infty
\quad \hbox{for any finite $a$.}
\end{equation}
The right-hand side of (6.6) is finite if $V_\Delta \in F_{n-1}^{(-)}$. 
Then, the right-hand side of (6.8) is also finite for $j \leq n-1$. So, we know from (5.26) that 
$\vert \bar r_n^\Delta \vert<\infty$ if $V_\Delta \in F^{(-)}_{n-1}$.
Since $\bar r_n^{\rm p}$ is always finite, it follows that
\begin{equation}
\vert \bar r_n \vert < \infty 
\quad
\hbox{if}
\quad 
V_\Delta \in F^{(-)}_{n-1}.
\end{equation}
If $\vert \bar r_N \vert <\infty$, then $\vert \bar r_n \vert <\infty$ for all $n \leq N$. Therefore, we can conclude that (3.8) makes sense if $V_\Delta \in F_{N-1}^{(-)}$.

\section{Small-$\boldsymbol{k}$ behavior of the remainder term}

In the previous section, it was shown that equation~(3.8) makes sense if $V_\Delta \in F^{(-)}_{N-1}$. 
In this section, we will show that $\bar \rho_N=o(k^N)$ as $k \to 0$ if $V_\Delta \in F^{(-)}_{N-1}$.

Using (3.5) and (5.17), we can write (3.10) as
\begin{equation}
\bar \rho_N
=(\rmi k)^{N+1} \int_{-\infty}^x
\bar Q(x,z;W;k) \,\bar q_N(x,\bar \omega(x,z;W;k))\,\rmd z,
\end{equation}
where
\begin{equation}
\bar Q(x,z;k;W)
\equiv
\frac{[\bar \tau(x,z;W;k)]^2}
{1-[\bar R_l(x,z;W;k)]^2}.
\end{equation}
Using (5.19), the right-hand side of (7.1) is split into two parts as
\begin{eqnarray}
\fl
\bar \rho_N
=(\rmi k)^{N+1} \int_{-\infty}^x
\bar Q(x,z;W;k) \,\bar q_N^{\rm p}(z,\bar \omega(x,z;W;k))\,\rmd z
\nonumber \\
+(\rmi k)^{N+1} \int_{-\infty}^x
\bar Q(x,z;W;k) \,\bar q_N^\Delta(z,\bar \omega(x,z;W;k))\,\rmd z.
\end{eqnarray}
Let $\bar \tau^{\rm p}$ and $\bar R_l^{\rm p}$ denote, respectively, $\bar \tau$ and $\bar R_l$ with $V$ replaced by $V_{\rm p}$.
We define
\begin{equation}
\bar Q_{\rm p}(x,z;k;W)
\equiv
\frac{[\bar \tau^{\rm p}(x,z;W;k)]^2}
{1-[\bar R_l^{\rm p}(x,z;W;k)]^2},
\end{equation}
\begin{equation}
\bar \omega_{\rm p}(x,z;W;k) \equiv V_{\rm p}(z)+\log \frac{1+\bar R_l^{\rm p}(x,z;W;k)}{1-\bar R_l^{\rm p}(x,z;W;k)}.
\end{equation}
In the limit $k\to 0$, we have the expressions
\begin{eqnarray}
\fl
\lim_{k \to 0} \bar \tau(x,z;W;k)={\rm sech}\,\frac{W-V(z)}{2},
\qquad
\lim_{k \to 0} \bar \tau^{\rm p}(x,z;W;k)={\rm sech}\,\frac{W-V_{\rm p}(z)}{2},
\nonumber \\
\fl
\lim_{k \to 0} \bar R_l(x,z;W;k)={\rm tanh}\,\frac{W-V(z)}{2},
\qquad
\lim_{k \to 0} \bar R_l^{\rm p}(x,z;W;k)={\rm tanh}\,\frac{W-V_{\rm p}(z)}{2}
\end{eqnarray}
(see equations~(2.18) of \cite{periodic}). Hence,
\numparts
\begin{eqnarray}
\lim_{k \to 0} \bar Q(x,z;W;k)=\lim_{k \to 0} \bar Q_{\rm p}(x,z;W;k)=1,
\\
\lim_{k \to 0} \bar \omega(x,z;W;k)=\lim_{k \to 0} \bar \omega_{\rm p}(x,z;W;k)=W.
\end{eqnarray}
\endnumparts
It can also be shown that
\begin{eqnarray}
\fl
\lim_{k \to 0}
\int_{-\infty}^x
\bar Q(x,z;W;k) \,\bar q_N^{\rm p}(z,\bar \omega(x,z;W;k))\,\rmd z
\nonumber \\
=\lim_{k \to 0}
\int_{-\infty}^x
\bar Q_{\rm p}(x,z;W;k) \,\bar q_N^{\rm p}(z,\bar \omega_{\rm p}(x,z;W;k))\,\rmd z,
\end{eqnarray}
provided that $V_\Delta \in F_0^{(-)}$.
(See appendix~A for the derivation.) 
It was shown in \cite{periodic} that
\begin{equation}
\lim_{k \to 0}
\int_{-\infty}^x
\bar Q_{\rm p}(x,z;W;k) \,\bar q_N^{\rm p}(z,\bar \omega_{\rm p}(x,z;W;k))\,\rmd z
=\mathcal{A}^{-1}_{\rm p}\bar q_N^{\rm p}(x,W).
\end{equation}
The right-hand side of (7.9) is equal to $\bar r_{N+1}^{\rm p}$ (see (5.24)). Therefore, 
\begin{equation}
\lim_{k \to 0}
\int_{-\infty}^x
\bar Q(x,z;W;k) \,\bar q_N^{\rm p}(x,\bar \omega(x,z;W;k))\,\rmd z =\bar r_{N+1}^{\rm p}(x,W).
\end{equation}
This gives the small-$k$ behavior of the first term of (7.3).

Next, we consider the second term of (7.3). 
Substituting (5.26) into (5.23), we write 
\begin{equation}
\bar q_N^\Delta= u + \sum_{m=1}^{N-1} v_m + w,
\end{equation}
\begin{equation}
\fl
u \equiv 2 \mathcal{B}(2 \mathcal{A}_0^{-1} \mathcal{B})^{N-1} \bar r_1^\Delta,
\quad \ \ 
v_m \equiv 2 \mathcal{B} (2 \mathcal{A}_0^{-1} \mathcal{B})^{m-1}\,
2 \mathcal{A}_0^{-1}\mathcal{B}_\Delta \,\bar r_{N-m}^{\rm p},
\quad \ \ 
w \equiv 2 \mathcal{B}_\Delta \,\bar r_N^{\rm p}.
\end{equation}
As shown in (7.7{\it b}), the quantity $\bar \omega(x,z;W;k)$ tends to $W$ as $k \to 0$.
This approach is uniform in $z$.
We fix $x$ and $W$, and let $k_0$ be a fixed (sufficiently small) positive number.
Using the same argument as in the last section, we can easily show that
\begin{eqnarray}
\left\vert
u(z,\bar \omega(x,z;W;k))
\right\vert
\leq
C h_N(z),
\qquad
\left\vert
v_m(z,\bar \omega(x,z;W;k))
\right\vert
\leq
C h_m(z),
\nonumber \\
\left\vert
w(z,\bar \omega(x,z;W;k))
\right\vert
\leq
C h_0(z)
\end{eqnarray}
for $-\infty<z<x$ and $\vert k \vert<k_0$, where we have defined
\begin{equation}
\fl
h_n(z) \equiv \int_{-\infty}^z(z-z')^{n-1} \left\vert V_\Delta(z') \right\vert \rmd z'
\quad (n\geq 1),
\qquad
h_0(z) \equiv \left\vert V_\Delta(z) \right\vert.
\end{equation}
It is easy to see that $\int_{-\infty}^x h_n(z)\, \rmd z < \infty$ if $V_\Delta \in F_n^{(-)}$. 
We have the inequality for $\bar Q$, 
\begin{equation}
\vert \bar Q(x,z;W;k) \vert \leq 1,
\end{equation}
as shown in appendix~B.
From (7.11), (7.13) and (7.15), we find that
\begin{equation}
\fl
\left\vert
\bar Q(x,z;W;k)\,
\bar q_N^\Delta(z,\bar \omega(x,z;W;k))
\right\vert
\leq
h(z),
\qquad
h(z) \equiv 
C \sum_{m=0}^N h_m(z).
\end{equation}
Obviously, $\int_{-\infty}^x h(z)\, \rmd z<\infty$ if $V_\Delta \in F_N^{(-)}$.
Therefore, by the dominated convergence theorem, the integral in the second term of (7.3) commutes with the limit $k \to 0$ if $V_\Delta \in F_N^{(-)}$. Namely, if $V_\Delta \in F_N^{(-)}$ we obtain, by using (7.7),
\begin{eqnarray}
\fl
\lim_{k \to 0}
\int_{-\infty}^x
\bar Q(x,z;W;k) \,\bar q_N^\Delta(x,\bar \omega(x,z;W;k))\,\rmd z
\nonumber \\
=\int_{-\infty}^x
\lim_{k \to 0}
\bar Q(x,z;W;k) \,\bar q_N^\Delta(x,\bar \omega(x,z;W;k))\,\rmd z
\nonumber \\
=\int_{-\infty}^x
\bar q_N^\Delta(x,W)\,\rmd z
= \mathcal{A}^{-1}_0\bar q_N^\Delta(x,W)
= \bar r_{N+1}^\Delta(x,W).
\end{eqnarray}

From (7.3), (7.10) and (7.17), we have
\begin{equation}
\lim_{k \to 0} \frac{1}{(\rmi k)^{N+1}} \bar \rho_N
=\bar r_{N+1}^{\rm p} + \bar r_{N+1}^\Delta = \bar r_{N+1}
\quad
\hbox{if}
\quad
V_\Delta \in F_N^{(-)}.
\end{equation}
Since $\bar \rho_N=(\rmi k)^{N+1} \bar r_{N+1} + \bar \rho_{N+1}$, 
equation~(7.18) implies that 
$ \lim_{k \to 0} \bar \rho_{N+1}/(\rmi k)^{N+1}=0$
if
$V_\Delta \in F_N^{(-)}$. 
Replacing $N$ by $N-1$, we obtain
\begin{equation}
\lim_{k \to 0}  \frac{1}{(\rmi k)^N} \bar \rho_N=0 
\quad
\hbox{if}
\quad
V_\Delta \in F_{N-1}^{(-)}.
\end{equation}
Therefore, if $V_\Delta \in F_{N-1}^{(-)}$, then $\bar R_r$ can be expanded to order $k^N$ as
\begin{equation}
\label{1-3.11}
\bar R_r(x,-\infty;W;k)=\bar r_0 + \rmi k \bar r_1 + (\rmi k)^2 \bar r_2 + \cdots + (\rmi k)^N \bar r_N+ o(k^N).
\end{equation}

\section{Expansion of $\boldsymbol{S}$}
It is shown in \cite{low} that the expansion of $S_r$ (defined by (2.7)) is obtained from (7.20) as
\begin{equation}
\fl
S_r(x,k)-\frac{1}{2}=a_0^{\rm R}(x) + \rmi k a_1^{\rm R}(x)+ (\rmi k)^2 a_2^{\rm R}(x)+ \cdots + (\rmi k)^N a_N^{\rm R}(x) + o(k^N),
\end{equation}
where 
\numparts
\begin{eqnarray}
a_0^{\rm R}(x)=\frac{1}{4} \lim_{W \to -\infty}
\rme^{-W+V(x)} \left[{\bar r}_0(x)-1\right],
\\
a_n^{\rm R}(x)=\frac{1}{4} \lim_{W \to -\infty}
\rme^{-W+V(x)} {\bar r}_n(x)
\qquad 
(n \geq 1).
\end{eqnarray}
\endnumparts
Equation (8.1) is valid if $V_\Delta \in F_{N-1}^{(-)}$, as is (7.20). 
Putting into (8.2) the expressions for $\bar r_0$, $\bar r_1$ and $\bar r_2$ (equations  (5.7), (5.14), (5.15), (5.27), (5.29)), we obtain
\numparts
\begin{eqnarray}
a_0^{\rm R}(x)=-\frac{1}{2} \rme^{V(x)-V_0},
\\
\fl
a_1^{\rm R}(x)
=\frac{1}{4L_0} \rme^{V(x)-V_0}
\biggl(
{}_{\rm p}[\hbox{$+$$-$}]_{x-L}^x-{}_{\rm p}[\hbox{$-$$+$}]_{x-L}^x
\biggr)
\nonumber \\
+\frac{1}{2} \rme^{V(x)-V_0}
\int_{-\infty}^x \rmd z
\left(
\rme^{V_0} \Delta^-(z) - \rme^{-V_0} \Delta^+(z)
\right),
\\
\fl
a_2^{\rm R}(x)
=\frac{1}{2L_0} \rme^{V(x)-V_0}
\left\{
\rme^{-V_0}  {}_{\rm p}[\hbox{$+$$-$$+$}]_{x-L}^x
-\frac{1}{8 L_0}
\left(
L_0^4 + 4 Q
\right)
\right\}
\nonumber \\
+\frac{1}{2L_0} \rme^{V(x)-2V_0}
\int_{-\infty}^x \rmd z\,
\Delta^+(z)
\biggl(
{}_{\rm p}[\hbox{$+$$-$}]_{z-L}^z-{}_{\rm p}[\hbox{$-$$+$}]_{z-L}^z
\biggr)
\nonumber \\
+ \rme^{V(x)-2 V_0}
\int_{-\infty}^x \rmd z
\left(
\rme^{V_0} \Delta^-(z) - \rme^{-V_0} \Delta^+(z)
\right)
[\hbox{$+$}]_z^x .
\end{eqnarray}
\endnumparts
The expansion of $S_l$ can be obtained in the same way.
Let $F^{(+)}_n$ denote the set of functions $v(x)$ satisfying
\begin{equation}
\int_a^\infty (1+ \vert x \vert^n)\vert v(x)\vert\, \rmd x < \infty
\quad \hbox{for any finite $a$.}
\end{equation}
If $V_\Delta \in F_{N-1}^{(+)}$, we have
\begin{equation}
\fl
S_l(x,k)-\frac{1}{2}=a_0^{\rm L}(x) + \rmi k a_1^{\rm L}(x)+ (\rmi k)^2 a_2^{\rm L}(x)+ \cdots + (\rmi k)^N a_N^{\rm L}(x) + o(k^N),
\end{equation}
where
\numparts
\begin{eqnarray}
a_0^{\rm L}(x)=-\frac{1}{2} \rme^{V(x)-V_0},
\\
\fl
a_1^{\rm L}(x)
=-\frac{1}{4L_0} \rme^{V(x)-V_0}
\biggl(
{}_{\rm p}[\hbox{$+$$-$}]_{x-L}^x-{}_{\rm p}[\hbox{$-$$+$}]_{x-L}^x
\biggr)
\nonumber \\
+\frac{1}{2} \rme^{V(x)-V_0}
\int_x^\infty\rmd z
\left(
\rme^{V_0} \Delta^-(z) - \rme^{-V_0} \Delta^+(z)
\right),
\\
\fl
a_2^{\rm L}(x)
=\frac{1}{2L_0} \rme^{V(x)-V_0}
\left\{
\rme^{-V_0}  {}_{\rm p}[\hbox{$+$$-$$+$}]_{x-L}^x
-\frac{1}{8 L_0}
\left(
L_0^4 + 4 Q
\right)
\right\}
\nonumber \\
-\frac{1}{2L_0} \rme^{V(x)-2V_0}
\int_x^\infty \rmd z\,
\Delta^+(z)
\biggl(
{}_{\rm p}[\hbox{$+$$-$}]_{z-L}^z-{}_{\rm p}[\hbox{$-$$+$}]_{z-L}^z
\biggr)
\nonumber \\
+ \rme^{V(x)-2 V_0}
\int_x^\infty \rmd z
\left(
\rme^{V_0} \Delta^-(z) - \rme^{-V_0} \Delta^+(z)
\right)
[\hbox{$+$}]_x^z.
\end{eqnarray}
\endnumparts

If $V_\Delta \in F_{N-1}^{(-)} \cap F_{N-1}^{(+)}$, i.e. if $V_\Delta \in L^1_{N-1}$, we can add together equations (8.1) and (8.5) to obtain the expansion of $S$ (equation (2.8)) as
\begin{equation}
\fl
S(x,k)-1=s_0(x) + \rmi k s_1(x) + (\rmi k)^2 s_2(x)
+\cdots + (\rmi k)^N s_N(x) + o(k^N)
\end{equation}
with 
\begin{equation}
s_n(x)=a_n^{\rm R}(x)+ a_n^{\rm L}(x).
\end{equation} 
From (8.3) and (8.6) we have
\numparts
\begin{eqnarray}
s_0(x)=-\rme^{V(x)-V_0},
\\
s_1(x)=
\frac{ \rme^{V(x)-V_0}}{2}
\int_{-\infty}^\infty \rmd z 
\left[
\rme^{V_0}\Delta^-(z)-\rme^{-V_0}\Delta^+(z)
\right],
\\
\fl
s_2(x)=
\frac{\rme^{V(x)-2V_0}}{2 L_0}
\Biggl\{
2\,{}_{\rm p}[\hbox{$+$$-$$+$}]_{x-L}^x
- \rme^{V_0}\left(\frac{L_0^3}{4}+ \frac{Q}{L_0}\right)
\nonumber \\
\qquad
+
\int_{-\infty}^x \rmd z 
\left({}_{\rm p}[\hbox{$+$$-$}]_{z-L}^z - {}_{\rm p}[\hbox{$-$$+$}]_{z-L}^z\right)
\Delta^+(z) 
\nonumber \\
\qquad
-
\int_x^{\infty} \rmd z
\left({}_{\rm p}[\hbox{$+$$-$}]_{z-L}^z - {}_{\rm p}[\hbox{$-$$+$}]_{z-L}^z\right)
\Delta^+(z) 
\nonumber \\
\qquad
+ 2 L_0
\int_{-\infty}^x \rmd z \,
\left[
\rme^{V_0}\Delta^-(z) -\rme^{-V_0}\Delta^+(z)
\right] 
[\hbox{$+$}]_z^x
\nonumber \\
\qquad
+ 2 L_0
\int_x^{\infty} \rmd z \,
\left[
\rme^{V_0}\Delta^-(z) -\rme^{-V_0}\Delta^+(z)
\right] 
[\hbox{$+$}]_x^z
\Biggr\}.
\end{eqnarray}
In a similar way (we omit the calculation), we can derive the expression for $s_3$ as
\begin{eqnarray}
\fl
s_3(x)
 =
\frac{\rme^{V(x)-2V_0}}{2L_0}
\Biggl\{
\int_{-\infty}^\infty \rmd z
\left(
3 \rme ^{-V_0} {}_{\rm p}[\hbox{$+$}\hbox{$-$}\hbox{$+$}]_{z-L}^z
+ \rme ^{V_0} {}_{\rm p}[\hbox{$-$}\hbox{$+$}\hbox{$-$}]_{z-L}^z
-\frac{L_0^3}{2}-\frac{2 Q}{L_0}
\right)
\Delta^+(z)
\nonumber \\
+
\int_{-\infty}^x \rmd z
\left(
{}_{\rm p}[\hbox{$+$}\hbox{$-$}]_{z-L}^z-{}_{\rm p}[\hbox{$-$}\hbox{$+$}]_{z-L}^z
\right)
\left[
3 \rme^{-V_0}\Delta^+(z)-\rme^{V_0}\Delta^-(z)
\right]
[\hbox{$+$}]_z^x
\nonumber \\
-
\int_x^\infty \rmd z
\left(
{}_{\rm p}[\hbox{$+$}\hbox{$-$}]_{z-L}^z-{}_{\rm p}[\hbox{$-$}\hbox{$+$}]_{z-L}^z
\right)
\left[
3 \rme^{-V_0}\Delta^+(z)-\rme^{V_0}\Delta^-(z)
\right]
[\hbox{$+$}]_x^z
\nonumber \\
+
2 L_0
\int_{-\infty}^x \rmd z
\left[
\rme^{V_0}\Delta^-(z) -\rme^{-V_0}\Delta^+(z)
\right]
\left(
3 \rme^{-V_0}[\hbox{$+$}\hbox{$+$}]_z^x-\rme^{V_0}[\hbox{$-$}\hbox{$+$}]_z^x
\right)
\nonumber \\
+
2 L_0
\int_x^\infty \rmd z
\left[
\rme^{V_0}\Delta^-(z) -\rme^{-V_0}\Delta^+(z)
\right]
\left(
3 \rme^{-V_0}[\hbox{$+$}\hbox{$+$}]_x^z-\rme^{V_0}[\hbox{$+$}\hbox{$-$}]_x^z
\right)
\Biggr\}.
\end{eqnarray}
\endnumparts

\section{Expansion of the Green function}
The expansion of the Green function in terms of $k$ can be obtained by substituting the power-series expression of $S$ into (2.9). 
To expand $G_{\rm S}$ to order $k^N$, we need the expansion of $S$ to order $k^{N+1}$ (as shown below).
Let us define
\begin{equation}
\fl
q_n(x,y)\equiv -\int_y^x  s_{n-1}(z) \,\rmd z,
\qquad
t_n(x) \equiv \frac{s_n(x)}{s_0(x)}=- \rme^{V_0-V(x)} s_n(x).
\end{equation}
If $V_\Delta \in L^1_N$, then equation~(8.7) holds with $N$ replaced by $N+1$. Substituting this into (2.9), we obtain equation~(1.10) with
\numparts
\begin{eqnarray}
g_{-1}(x,y)=\frac{1}{2 \sqrt{s_0(x)s_0(y)}},
\\
g_0(x,y)
=\left[q_1(x,y)-\frac{t_1(x)}{2}-\frac{t_1(y)}{2}\right] g_{-1}(x,y),
\\
\fl
g_1(x,y)=
\Biggl\{
q_2(x,y)+\frac{[q_1(x,y)]^2}{2}  -\frac{q_1(x,y)}{2}[t_1(x)+t_1(y)] 
\nonumber \\
-\frac{t_2(x)}{2}
-\frac{t_2(y)}{2}
+\frac{3[t_1(x)]^2}{8} + \frac{3[t_1(y)]^2}{8}
+\frac{t_1(x) t_1(y)}{4} 
\Biggr\}\, g_{-1}(x,y),
\\
\fl
g_2(x,y)=\Biggl\{
q_3(x,y)+q_1(x,y) q_2(x,y)+\frac{[q_1(x,y)]^3}{6} \nonumber \\
-\left(\frac{[q_1(x,y)]^2}{4} + \frac{q_2(x,y) }{2} \right)\left[t_1(x)+t_1(y)\right]
\nonumber \\
+q_1(x,y)
\left(
\frac{3[t_1(x)]^2}{8} + \frac{3[t_1(y)]^2}{8} + \frac{t_1(x)t_1(y)}{4}
-\frac{t_2(x)}{2}-\frac{t_2(y)}{2}
\right)
\nonumber \\
-\frac{t_3(x)}{2}-\frac{t_3(y)}{2}
+\frac{3 t_1(x)t_2(x)}{4}+\frac{3 t_1(y)t_2(y)}{4}
-\frac{5[t_1(x)]^3}{16}-\frac{5[t_1(y)]^3}{16}
\nonumber \\
+\frac{t_2(x)t_1(y)}{4}+\frac{t_2(y)t_1(x)}{4}
-\frac{3 t_1(x)t_1(y)}{16}\left[t_1(x)+t_1(y)\right]
\Biggr\}\, g_{-1}(x,y),
\end{eqnarray}
\endnumparts
and so on. 
It is easy to see that we need $s_0, s_1, \ldots, s_{N+1}$ in order to calculate $g_N$.  
Thus, (1.10) holds if $V_\Delta \in L^1_N$.
From (8.9) and (9.2), the explicit forms of $g_{-1}$ and $g_0$ are obtained as
\numparts
\begin{eqnarray}
g_{-1}(x,y)=\frac{\rme^{V_0}}{2}\,\rme^{-[V(x)+V(y)]/2},
\\
\fl
g_0(x,y)=\frac{1}{2}\,
\Biggl\{
\,[\hbox{$+$}]_y^x +\frac{ \rme^{V_0}}{2}
\int_{-\infty}^\infty \rmd z 
\left[
\rme^{V_0}\Delta^-(z)-\rme^{-V_0}\Delta^+(z)
\right]
\Biggr\}\,\rme^{-[V(x)+V(y)]/2}.
\end{eqnarray}
\endnumparts

\section{More general potentials}

The method presented in this paper is also applicable to the cases where $V(x)$ does not approach the same periodic function as $x \to -\infty$ and $x \to +\infty$. 
Suppose that $V(x) \to V_{{\rm p}1}(x)$ as $x \to -\infty$ and $V(x) \to V_{{\rm p}2}(x)$ as $x \to +\infty$, where $V_{{\rm p}1}$ and $V_{{\rm p}2}$ are two different periodic functions with periods $L_1$ and $L_2$ respectively. Namely,
\begin{eqnarray}
\fl
V=V_{{\rm p} 1}+V_{\Delta 1}=V_{{\rm p} 2}+V_{\Delta 2},
\qquad
V_{{\rm p} 1}(x+L_1)=V_{{\rm p} 1}(x),
\qquad
V_{{\rm p} 2}(x+L_2)=V_{{\rm p} 2}(x),
\nonumber \\
\lim_{x \to -\infty} V_{\Delta 1}(x)=0,
\qquad
\lim_{x \to +\infty} V_{\Delta 2}(x)=0.
\end{eqnarray}
The expressions for $a^{\rm R}_n$ and $a^{\rm L}_n$ given in section~8 (equations (8.3) and (8.6)) still hold if we use $V_{{\rm p}1}$ for $a^{\rm R}_n$ and $V_{{\rm p}2}$ for $a^{\rm L}_n$. 
The expansion of $G_{\rm S}$ is obtained by calculating $s_n=a^{\rm R}_n+a^{\rm L}_n$, and substituting into (9.1) and (9.2). 

Let ${}_{{\rm p} i}[\sigma_1,\ldots,\sigma_n]_a^b$, $V_{0i}$, $L_{0i}$ and $\Delta^\pm_i$ ($i=1,2$) be the quantities defined in the same way as ${}_{\rm p}[\sigma_1,\ldots,\sigma_n]_a^b$, $V_0$, $L_0$ and $\Delta^\pm$, respectively, with $V_{\rm p}$, $V_\Delta$ and $L$ replaced by $V_{{\rm p}i}$, $V_{\Delta i}$ and $L_i$. 
If $V_{\Delta 1} \in F^{(-)}_N$ and $V_{\Delta 2} \in F^{(+)}_N$, the Green function can be expanded to order $k^N$ in the form of (1.10). Instead of (9.3), we have
\numparts
\begin{eqnarray}
g_{-1}(x,y)=\frac{1}{\rme^{-V_{01}}+\rme^{-V_{02}}}\,\rme^{-[V(x)+V(y)]/2},
\\
g_0(x,y)=\frac{1}{2}\,
\Biggl\{
\,[\hbox{$+$}]_y^x 
-\frac{t_1(x)+t_1(y)}{\rme^{-V_{01}}+\rme^{-V_{02}}}
\Biggr\}\,\rme^{-[V(x)+V(y)]/2},
\end{eqnarray}
\endnumparts
where
\begin{eqnarray}
\fl
t_1(x)
=\frac{-1}{\rme^{-V_{01}}+\rme^{-V_{02}}}
\Biggl\{
\frac{\rme^{-V_{01}}}{2L_{01}} \biggl(
{}_{{\rm p}1}[\hbox{$+$$-$}]_{x-L_1}^x-{}_{{\rm p}1}[\hbox{$-$$+$}]_{x-L_1}^x
\biggr)
\nonumber \\
\fl
\qquad \qquad
-
\frac{\rme^{-V_{02}}}{2L_{02}} \biggl(
{}_{{\rm p}2}[\hbox{$+$$-$}]_{x-L_2}^x-{}_{{\rm p}2}[\hbox{$-$$+$}]_{x-L_2}^x
\biggr)
+\rme^{-V_{01}}
\int_{-\infty}^x \rmd z
\left[
\rme^{V_{01}} \Delta_1^-(z) - \rme^{-V_{01}} \Delta_1^+(z)
\right]
\nonumber \\
+\rme^{-V_{02}}
\int_x^\infty\rmd z
\left[
\rme^{V_{02}} \Delta_2^-(z) - \rme^{-V_{02}} \Delta_2^+(z)
\right]
\Biggr\}.
\end{eqnarray}

It is equally easy to calculate the expansion of the Green function for the cases in which $V(x)$ is asymptotically periodic as $x \to -\infty$ and $V(x) \to \pm \infty$ as $x \to +\infty$. For such cases, we can use (8.3) together with the expressions for $a^{\rm L}_n$ given in \cite{low}.

\section{Generic case for the Schr\"odinger equation}

Let us suppose that the Schr\"odinger equation is given with an asymptotically periodic $V_{\rm S}$ satisfying (1.12), and that the corresponding Fokker-Planck potential $V$ is yet unknown. 
We assume that there are no bound states, and we shift the origin of the energy scale ($k=0$) to the bottom of the lowest band. The Schr\"odinger equation at $k=0$ reads
\begin{equation}
-\frac{\rmd^2}{\rmd x^2}\psi_0(x)+V_{\rm S}(x)\psi_0(x)=0.
\end{equation}
Let  $\psi_0^+(x)$ and $\psi_0^-(x)$ be the solutions of (11.1) which remain bounded as $x \to +\infty$ and $x \to -\infty$, respectively. 
We can choose the phase of $\psi_0^\pm$ so that $\psi_0^\pm(x)>0$ for any $x$. 
(Although $\psi_0^\pm$ still has an arbitrariness of a positive multiplication factor, this arbitrariness does not remain in the final result for the Green function.) 
Let us define
\begin{equation}
V^\pm(x)\equiv -2 \log \psi_0^\pm(x), 
\qquad
f^\pm(x)\equiv -\frac{1}{2}\frac{\rmd}{\rmd x} V^\pm(x).
\end{equation}
It is easy to check that both $f^+$ and $f^-$ satisfy (1.4). So both $V^+$ and $V^-$ are Fokker-Planck potentials corresponding to $V_{\rm S}$. 
Since $V^+(x)$ and $V^-(x)$ tend to periodic functions as $x \to +\infty$ and $x \to -\infty$, respectively, we can write $V^\pm = V^\pm_{\rm p} + V^\pm_\Delta$, where
\begin{equation}
\fl
V^\pm_{\rm p}(x+L)= V^\pm_{\rm p}(x)
\qquad
\lim_{x\to +\infty} V^+_\Delta(x)=0, \qquad \lim_{x\to -\infty} V^-_\Delta(x)=0.
\end{equation}

If $\psi_0^+$ and $\psi_0^-$ coincide up to a multiplicative factor, then $V^+=V^- +C$ with a constant $C$, and so $V^\pm(x)$ are asymptotically periodic for both $x\to + \infty$ and $x \to -\infty$ (that is, we have $\lim_{x \to -\infty}V^+_\Delta(x)=0$ and $\lim_{x \to +\infty}V^-_\Delta(x)=0$ in addition to (11.3)).  This is the so-called exceptional case. In this exceptional case, we can use the result of section~9 for the Green function, using either $V^+$ or $V^-$ in place of $V$. On the other hand, if $\psi_0^+$ and $\psi_0^-$ are linearly independent, then $V^+(x)$ and $V^-(x)$ are not asymptotically periodic for $x \to -\infty$ and $x \to +\infty$, respectively. This is the ^^ generic case'.  

To deal with the generic case, we can use the method described in section~7 of \cite{low}. As explained there, the expansion of $S$ takes the form
\begin{eqnarray}
\fl
S(x,k)-1=(\rmi k)^{-1} s_{-1}(x)  +s_0(x) + \rmi k s_1(x) + (\rmi k)^2 s_2(x)
\nonumber \\
+\cdots + (\rmi k)^N s_N(x) + o(k^N)
\end{eqnarray}
with
\begin{equation}
s_{-1}=\frac{1}{2} \left[f^-(x) - f^+(x) \right],
\qquad
s_n=a_n^{{\rm R} -} + a_n^{{\rm L}+} \quad (n \geq 0),
\end{equation}
where $a_n^{\rm R\pm}$ and $a_n^{\rm L\pm}$ denote, respectively, $a_n^{\rm R}$ and $a_n^{\rm L}$ with $V$ replaced by $V^\pm$. 
Equation (11.4) is valid if $V_\Delta^- \in F_{N-1}^{(-)}$ and $V_\Delta^+ \in F_{N-1}^{(+)}$.
This is a generalization of (8.7). 
In the exceptional case (where $V^+=V^- + C$), we have $a_n^{\rm R +}=a_n^{\rm R -}$, $a_n^{\rm L +}=a_n^{\rm L -}$ and $s_{-1}=0$, and  hence (11.4) reduces to (8.7). 
If $\psi_0^\pm$ are linearly independent, then $f^+(x) \neq f^-(x)$ for any $x$, as can be easily verified. 
So, in the generic case, $s_{-1}(x) \neq 0$ for any $x$.

In the generic case, the expansion of $G_{\rm S}$ to order $k^N$ is obtained from the expansion of $S$ to order $k^{N-1}$. 
Substituting (11.4) (with $N$ replaced by $N-1$) into (2.9), we have
\begin{eqnarray}
\fl
G_{\rm S}(x,y;k)=g_0(x,y)+\rmi k g_1(x,y)+ (\rmi k)^2 g_2(x,y)
+ \cdots + (\rmi k)^N g_N(x,y) + o(k^N),
\end{eqnarray}
where
\numparts
\begin{eqnarray}
g_0(x,y)=\frac{-\exp[q_0(x,y)]}{2 \sqrt{s_{-1}(x)s_{-1}(y)}},
\\
g_1(x,y)
=\left[q_1(x,y)-\frac{s_0(x)}{2s_{-1}(x)}-\frac{s_0(y)}{2s_{-1}(y)}\right] g_0(x,y),
\\
\fl
g_2(x,y)=
\Biggl\{
q_2(x,y)+\frac{[q_1(x,y)]^2}{2}  
-\frac{q_1(x,y)}{2}\left[\frac{s_0(x)}{s_{-1}(x)}+\frac{s_0(y)}{s_{-1}(y)}\right]
-\frac{1}{2}\left[
\frac{s_1(x)}{s_{-1}(x)}
+\frac{s_1(y)}{s_{-1}(y)}
\right]
\nonumber \\
+\frac{3}{8}\left[\frac{s_0(x)}{s_{-1}(x)}\right]^2 
+ \frac{3}{8}\left[\frac{s_0(y)}{s_{-1}(y)}\right]^2
+\frac{s_0(x) s_0(y)}{4 s_{-1}(x)s_{-1}(y)} 
\Biggr\}\, g_0(x,y),
\end{eqnarray}
\endnumparts
and so on, with $q_n$ defined by (9.1). 
For $N \geq 2$, equation~(11.6) is valid if $V_\Delta^- \in F_{N-2}^{(-)}$ and $V_\Delta^+ \in F_{N-2}^{(+)}$. 

The explicit forms of $g_0$ and $g_1$ obtained from (11.7), (11.5), (8.3) and (8.6) are
\numparts
\begin{eqnarray}
g_0(x,y)=\frac{-\exp\left\{\left[V^-(x)-V^+(x)-V^-(y)+V^+(y)\right]/4\right\}}
{\sqrt{\left[f^-(x)-f^+(x)\right]\left[f^-(y)-f^+(y)\right]}},
\\
\fl
g_1(x,y)=
\frac{1}{2}
\Biggl[
\int_y^x \left(\rme^{V^-(z)-V_0^-} + \rme^{V^+(z)-V_0^+}\right)\, \rmd z
\nonumber \\
+ \frac{\rme^{V^-(x)-V_0^-} + \rme^{V^+(x)-V_0^+}}{f^-(x) - f^+(x)}
+ \frac{\rme^{V^-(y)-V_0^-} + \rme^{V^+(y)-V_0^+}}{f^-(y) - f^+(y)}
\Bigg]g_0(x,y),
\end{eqnarray}
\endnumparts
where $V_0^+$ and $V_0^-$ are defined by (2.12) with $V_{\rm p}$ replaced by $V_{\rm p}^+$ and $V_{\rm p}^-$, respectively. 
By substituting (11.2) into (11.8{\it a}), we can reproduce the well-known expression for $g_0$:
\begin{equation}
g_0(x,y)=-\frac{\psi_0^+(x) \psi_0^-(y)}{W[\psi_0^+, \psi_0^-]},
\end{equation}
where $W$ denotes the Wronskian defined by $W[\psi,\phi] \equiv \psi \phi'- \psi' \phi$.

\section{Examples}
Let us demonstrate the calculation of the expansion with simple examples.

\bigskip
\noindent
{\bf Example 1.}

\nobreak
\noindent
We consider the potential $V(x)=V_{\rm p}(x) +V_\Delta(x)$ with
\numparts
\begin{equation}
V_{\rm p}(x)=
\cases{
0 & $(0<x<a)$ \\
C & $(a<x<L)$ \\
},
\qquad 
V_{\rm p}(x+L)=V_{\rm p}(x),
\end{equation}
\begin{equation}
V_\Delta(x)=
\cases{
-h & $(0<x<a)$ \\
0 & (otherwise) \\
}.
\end{equation}
\endnumparts
This potential is illustrated in figure~1.
%
%
%
\begin{figure}
\hspace{1cm}
\includegraphics[scale=0.5]{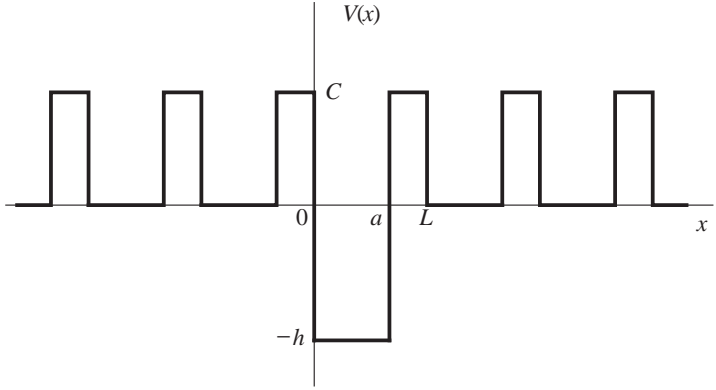}
\caption{
The potential $V(x)$ of example~1 (equations~(12.1)).
}
\end{figure}
The Green function can be exactly obtained for this potential.
Before showing the calculation of the low-energy expansion, let us first explain about the exact Green function and its properties.

For any potential, the Green function can be expressed in terms of $\tau$, $R_r$ and $R_l$ as
\begin{equation}
\fl
G_{\rm S}(x,y;k)
=\frac{[1+R_l(\infty,x;k)][1+R_r(y,-\infty;k)]\, \tau(x,y;k)}
{2 \rmi k[1-R_l(\infty,x;k)R_r(x,-\infty;k)][1-R_l(x,y;k)R_r(y,-\infty;k)]}
\end{equation}
(see equation~(3.6a) of \cite{expressions}).  
We confine ourselves to the case $0<y\leq x < a$. 
Then, for this potential, $R_l(x,y;k)=0$ and $\tau(x,y;k)=\rme^{\rmi k(x-y)}$.  So, (12.2) becomes
\begin{equation}
G_{\rm S}(x,y;k)
=\frac{[1+R_l(\infty,x;k)][1+R_r(y,-\infty;k)]\, \rme^{\rmi k(x-y)}}
{2 \rmi k[1-R_l(\infty,x;k)R_r(x,-\infty;k)]}.
\end{equation}
The exact expressions of the reflection coefficients are
\begin{equation}
\fl
R_r(x,-\infty;k)=
\rme^{2 \rmi k x}\frac{\tanh(h/2)+R_0}{1+R_0 \tanh(h/2)},
\qquad
R_l(\infty,x;k)=R_r(a-x,-\infty;k)
\end{equation}
for $0<x<a$, where
\begin{eqnarray}
\fl
\alpha(k) \equiv 
\frac{\rme^{-\rmi k L}}{1-A^2}\left(1-A^2 \rme^{2 \rmi k b}\right),
\qquad
\beta(k) \equiv 
\frac{\rme^{-\rmi k L}}{1-A^2} A \left(\rme^{2 \rmi k b}-1\right),
\qquad
A \equiv \tanh \frac{-C}{2},
\nonumber \\
R_0 \equiv
\frac{-\alpha(k)+\alpha(-k) - \rmi \sqrt{4-[\alpha(k)+\alpha(-k)]^2}}{2\beta(-k)}.
\end{eqnarray}
(We omit the derivation of these expressions.)
The exact $G_{\rm S}$ is given by (12.3) with (12.4).
The graphs of the exact ${\rm Re}\,G_{\rm S}$ and ${\rm Im}\,G_{\rm S}$, as functions of real $k$ with fixed $x$ and $y$, are shown in figures~2 and~3.
%
%
%
\begin{figure}
\hspace{1cm}
\includegraphics[scale=0.75]{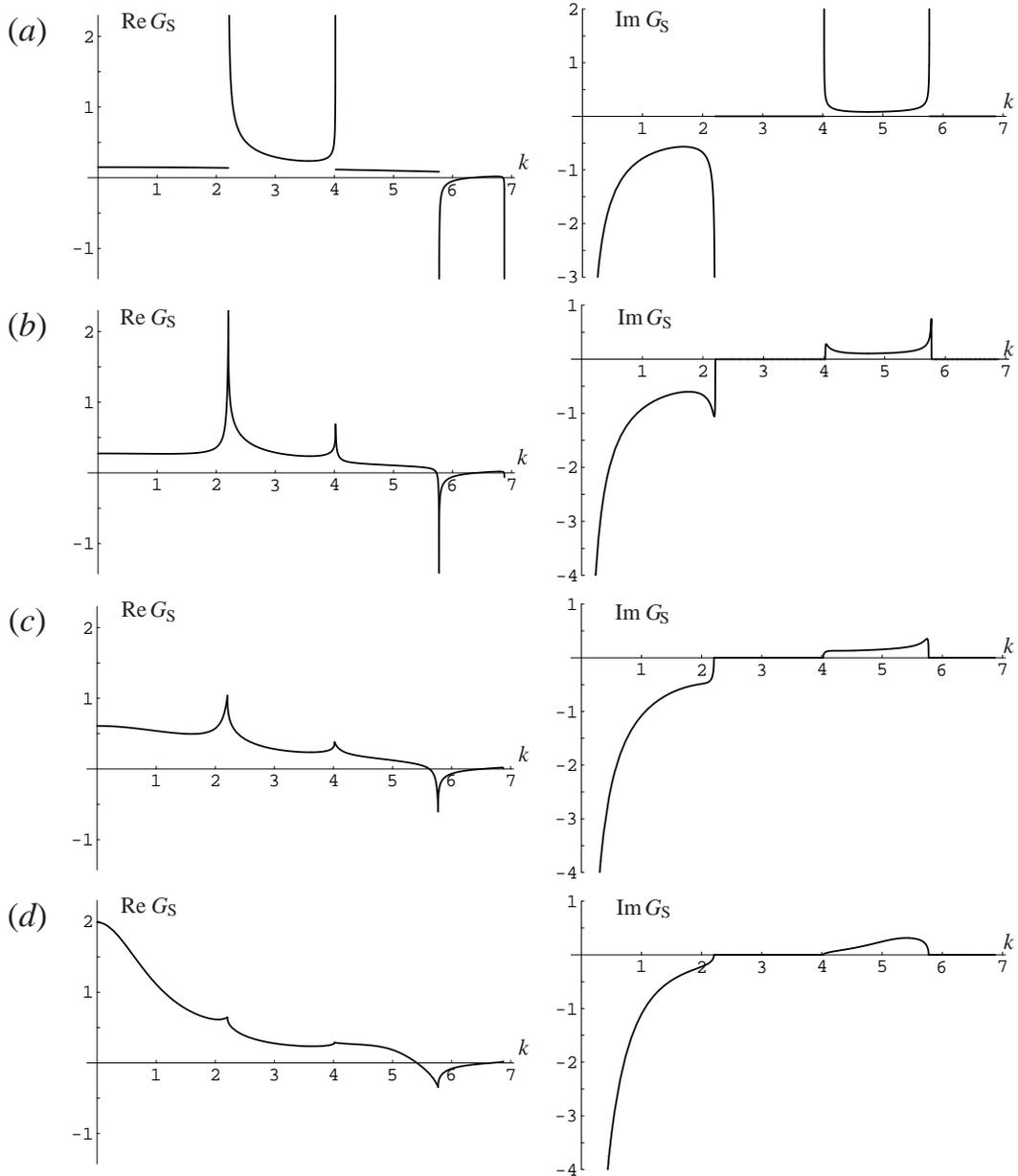}
\caption{
The real and imaginary parts of $G_{\rm S}(x,y;k)$ for the potential shown in figure~1, plotted as functions of real $k$, with various values of $h$.
(a)~$h=0$, (b)~$h=0.2$, (c)~$h=0.5$, (d)~$h=1$. 
In all the graphs, $C=1$, $L=1$, $a=0.6$, $x=0.4$ and $y=0.1$.
The graphs are shown here for the range of $k$ in the lowest two energy bands and the lowest two gaps. The bands are $0<k<k_1$ and $k_2<k<k_3$, and the gaps are $k_1<k<k_2$ and $k_3<k<k_4$, where $k_1 \simeq 2.21$, $k_2 \simeq 4.02$, $k_3 \simeq 5.77$ and $k_4 \simeq 6.88$. 
The imaginary part of $G_{\rm S}$ is identically zero in the gaps. 
}
\end{figure}
%
%
%
%
%
%
%
\begin{figure}
\hspace{1cm}
\includegraphics[scale=0.75]{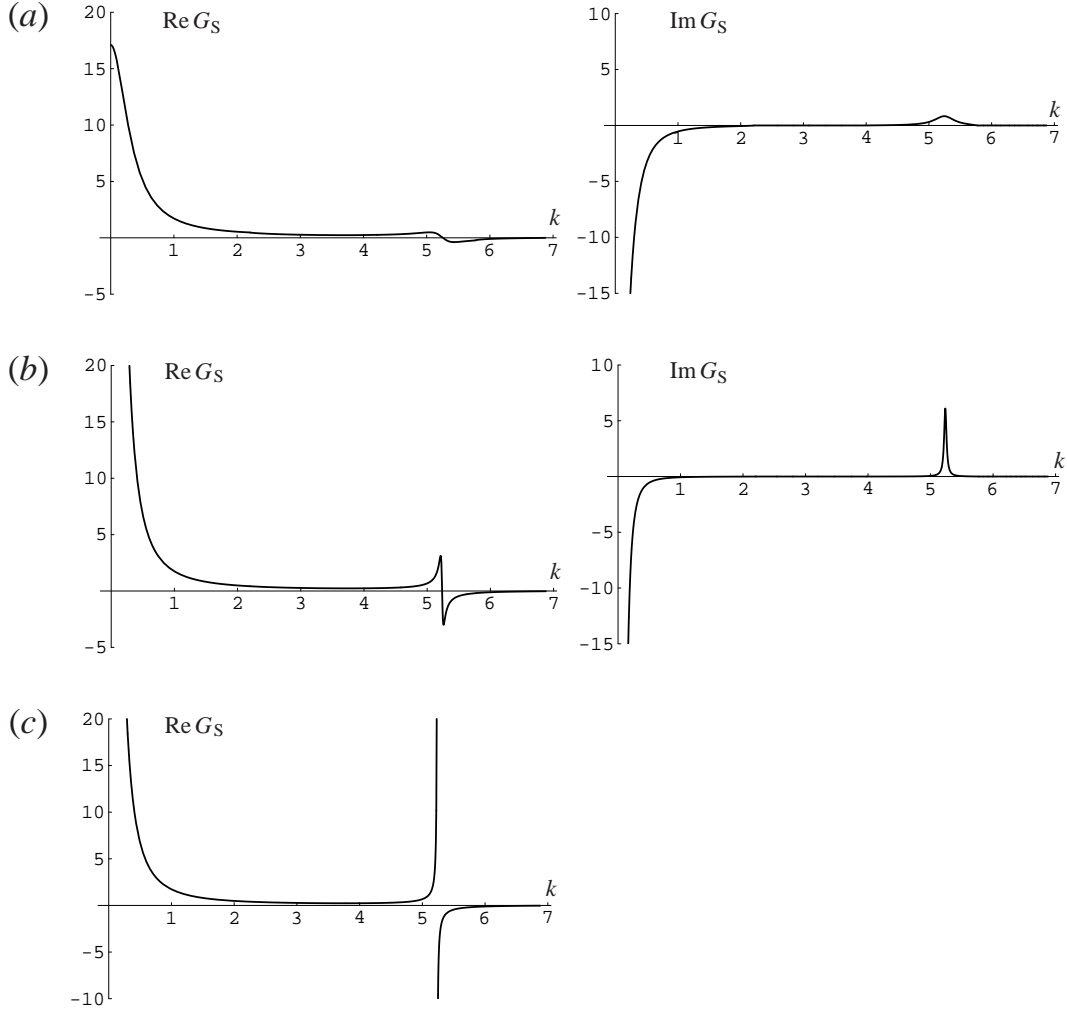}
\caption{
Same as figure~2, with larger values of $h$. (a) $h=2$, (b) $h=4$, (c) $h=\infty$.
In (b), the real part of $G_{\rm S}(k)$ tends to a finite value ($\simeq 1000$) as $k \to 0$, although the graph is truncated at the top.
The graph in (c) is the plot of (12.6). This graph diverges to infinity as $k \to 0$. 
The singularity at $k=\pi/a\simeq 5.24$ corresponds to an eigenvalue for the infinite square well potential. 
The peak around $k=\pi/a$ in the graphs of ${\rm Im}\,G_{\rm S}$ becomes a delta function as we let $h \to +\infty$.
}
\end{figure}
The case $h=0$ corresponds to a purely periodic potential, which was studied section~11 of~\cite{periodic}. When $h=0$, the graphs are discontinuous at the edges of the energy bands (figure~2(a)). 
The graphs become continuous when $h\neq 0$, but singularities remain at the edges of the bands (figures~2 (b)--(d)). As $h$ is further increased, these singularities become less prominent in the graphs (figures~3(a) and (b)).
In the extreme case $h=+\infty$, the potential is an infinite square well, for which the exact Green function is
\begin{equation}
G_{\rm S}(x,y;k)=
\frac{\cos[k(x-a)] \cos (k y)}
{k \sin (k a)}
\end{equation}
(see figure~3(c)).
As $h$ becomes large, the Green function approaches (12.6) except at $k= n \pi/a$ ($n$ integer), where the right-hand side (12.6) has poles.

Now let us turn to the low-energy expansion.
Since $V_\Delta \in L^1_N$ for any $N$, expansion (1.10) is valid for any $N$.
We need to calculate the various quantities appearing in (8.9).
The quantities involving only $V_{\rm p}$ (and not $V_\Delta$) have already been calculated in \cite{periodic}. 
It is shown in section~11 of \cite{periodic} that
\numparts
\begin{eqnarray}
L_0=\sqrt{(a+b \rme^{C})(a + b \rme^{-C})},
\qquad
V_0=\frac{1}{2}\log \frac{a+b \rme^{C}}{a + b \rme^{-C}},
\\
Q=\frac{1}{12} (a^4 + 6 a^2 b^2 + b^4) + \frac{a b}{3}(a^2 + b^2) \cosh C,
\end{eqnarray}
\endnumparts
\numparts
\begin{eqnarray}
{}_{\rm p}[\hbox{$+$$-$}]_{x-L}^x=\frac{1}{2}(a^2+b^2)+ \rme^{-C} b (a-x) +\rme^{C} b x,
\\
{}_{\rm p}[\hbox{$+$$-$$+$}]_{x-L}^x=
\frac{a}{6}(a^2+3b^2)+ 2 b x (x-a) \sinh C + \frac{b}{6} (3 a^2 + b^2) \rme^C.
\end{eqnarray}
\endnumparts
The expressions for ${}_{\rm p}[\hbox{$-$$+$}]_{x-L}^x$ and ${}_{\rm p}[\hbox{$-$$+$$-$}]_{x-L}^x$ are obtained from (12.8{\it a}) and (12.8{\it b}), respectively, by changing the sign of $C$.
Other integrals in (8.9) are also easy to calculate. 
For $0<z\leq x <a$, we can see that
\begin{equation}
\fl
[\hbox{$+$}]_z^x= \rme^{-h} (x-z),
\qquad
[\hbox{$+$$+$}]_z^x= \frac{\rme^{-2h}}{2} (x-z)^2,
\qquad
[\hbox{$-$$+$}]_z^x=[\hbox{$+$$-$}]_z^x=\frac{1}{2} (x-z)^2.
\end{equation}
The functions $\Delta^\pm$ defined by (5.16) are
\begin{equation}
\Delta^{\pm}(z)=
\cases{
\rme^{\mp h}-1 & $(0<z<a)$  \\
0 & (otherwise)
}.
\end{equation}
The integrals including $\Delta^\pm$ can be calculated, for example, as
\begin{equation}
\int_{-\infty}^\infty \rmd z \left[ \rme^{V_0} \Delta^-(z)-\rme^{-V_0} \Delta^+(z)\right]
=2 a \left[\sinh (V_0+h) -\sinh V_0 \right].
\end{equation}
Substituting the above expressions in (8.9), we obtain
\begin{eqnarray}
\fl
s_0=-\rme^{-(V_0+h)},
\qquad
s_1=\rme^{-(V_0+h)} a [\sinh (V_0+h)- \sinh V_0],
\nonumber \\
\fl
s_2=
\frac{\rme^{-(2V_0+h)}}{L_0}
\Biggl[
\frac{a}{6}(a^2+3 b^2) +\frac{b}{6}(3 a^2 + b^2)\rme^C
+ 2 b x(x- a) \rme^{-h} \sinh C 
-\frac{\rme^{V_0}}{8 L_0}\left(L_0^4+ 4Q\right)
\Biggr]
\nonumber \\
+\rme^{-2(V_0 + h)}\left[x^2 + (a-x)^2 \right][\sinh (V_0+h)-\sinh V_0],
\nonumber \\
\fl
s_3=\frac{\rme^{-(2 V_0+h)}}{2 L_0}
a \left(\rme^{-h}-1 \right)
\Biggl[
\left(3 \rme^{-V_0} + \rme^{V_0}\right) \frac{a}{6}(a^2 + 3 b^2)
+\left(3 \rme^{-V_0+C} + \rme^{V_0-C}\right) \frac{b}{6}(3 a^2 + b^2)
\nonumber \\
-b a^3 \left(\rme^{-h} + 1\right) \rme^{-V_0} \sinh C 
-\frac{1}{2L_0}\left(L_0^4+4Q\right)
\Biggr]
\nonumber \\
+ \frac{\rme^{-(2 V_0 + h)}}{3}
\left[ x^3 - (x-a)^3 \right]
\left(
3 \rme^{-V_0 - 2 h} - \rme^{V_0}
\right)
\left[\sinh (V_0+h)-\sinh V_0 \right],
\nonumber \\
\end{eqnarray}
with $L_0$, $V_0$ and $Q$ given by (12.7). (We have used $L_0=b \sinh C/\sinh V_0$.)
Substituting (12.12) into (9.1), and then into (9.2), we obtain $g_{-1}$, $g_0$, $g_1$ and $g_2$. 
In particular,
\begin{equation}
\fl
g_{-1}=\frac{\rme^{V_0+h}}{2},
\qquad
g_0=\frac{1}{2}(x-y) + \frac{a}{2}
\,\rme^{V_0+h}\left[\sinh(V_0+h) - \sinh V_0 \right],
\end{equation}
as can also be seen directly from (9.3).
Thus, the expansion of the Green function is obtained to order $k^2$.
This result is shown in figure~4 (the broken lines). 
%
%
%
\begin{figure}
\hspace{1cm}
\includegraphics[scale=0.75]{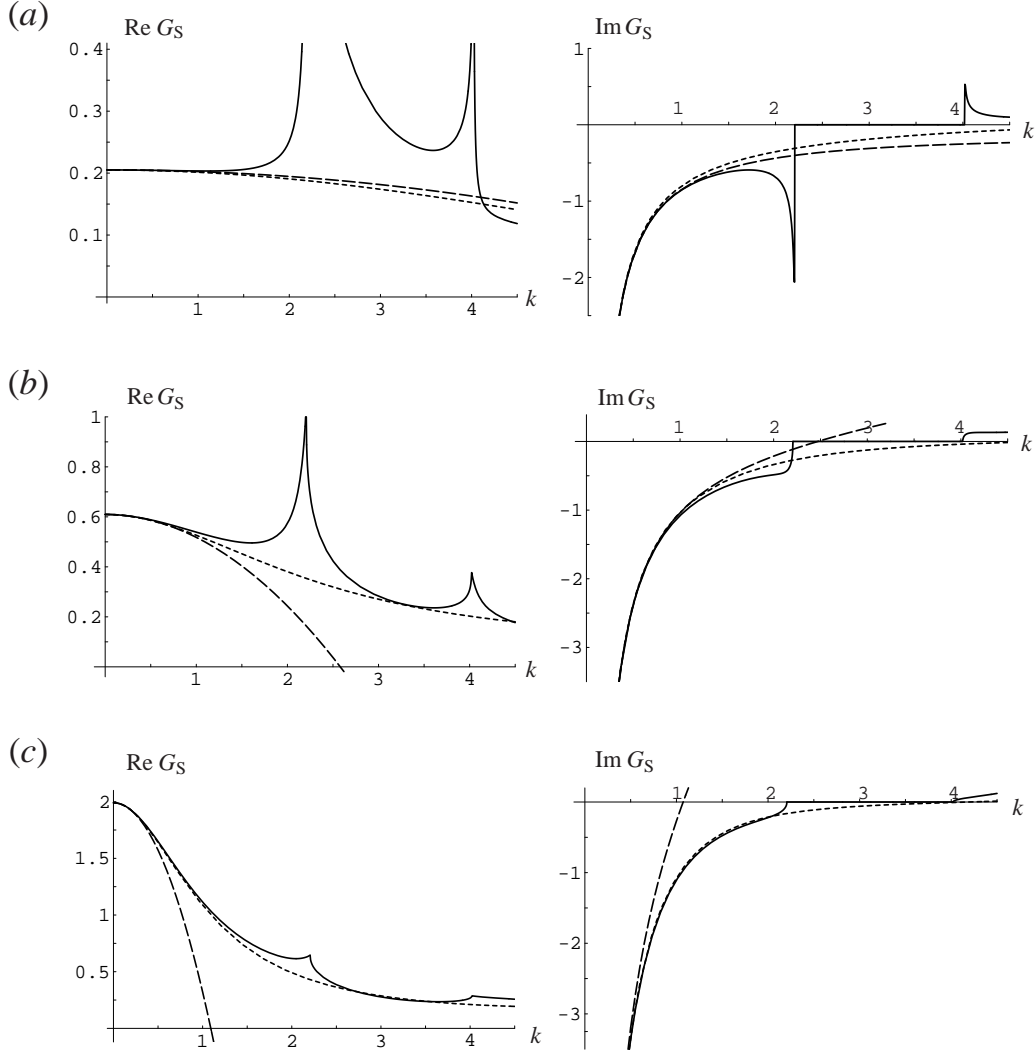}
\caption{
${\rm Re}\,G_{\rm S}(x,y;k)$ and ${\rm Im}\,G_{\rm S}(x,y;k)$ plotted as functions of real $k$, for (a) $h=0.1$, (b) $h=0.5$ and (c) $h=1$. (All the other parameters are the same as in figures~2 and 3.) 
The solid lines show the exact Green function (close-up of figure~2).
The broken lines show the result of the low-energy expansion up to order $k^2$.
The dotted lines represent the approximation given by (12.21).
}
\end{figure}
We can see from these graphs that this expansion is indeed correct. 

For a potential like (12.1), there is a good approximation method which can be used for a wide range of $k$. Let us explain this approximation in connection with the power-series expansion of $\bar R_r$. 
From (5.7), (5.14), (5.15), (5.27) and (5.29), we have
\numparts
\begin{eqnarray}
\fl
\bar r_0=\tanh \frac{V_0-W}{2},
\qquad
\bar r_1=\frac{1}{\cosh^2 \frac{W-V_0}{2}}
\left[
x \sinh (V_0+h)  - \frac{a}{2} \sinh V_0
\right],
\\\fl
\bar r_2 =\frac{1}{\cosh^3 \frac{W-V_0}{2}}
\Biggl\{
\left[
x^2  \sinh (V_0+h)  - a x \sinh V_0
\right]\cosh \frac{W+V_0+2h}{2}
\nonumber \\
-\frac{a}{12 L_0}(a^2+3 b^2) \sinh \frac{W+V_0}{2} 
-\frac{b}{12 L_0}(3a^2+ b^2) \sinh \frac{W+V_0-2C}{2}
\nonumber \\
+
\frac{1}{16L_0^2}\left(L_0^4+4Q\right) \sinh \frac{W-V_0}{2}
\Biggr\}.
\end{eqnarray}
\endnumparts
Recall that $R_r(x,-\infty;k)$ is obtained from $\bar R_r(x,-\infty;W;k)$ by setting $W=V(x)$. Since $V(x)=-h$ for $0<x<a$, the expansion of $R_r(x,-\infty;k)$ for $0<x<a$ is 
\begin{equation}
R_r(x,-\infty;k)=r_0 + \rmi k r_1 + (\rmi k)^2 r_2 + \cdots,
\end{equation}
where $r_0$ and $r_1$ are obtained by letting $W=-h$ in (12.14{\it a}) as
\begin{equation}
r_0= \tanh \frac{V_0+h}{2}, \qquad
r_1= 2(x- \delta)r_0,
\end{equation}
with $\delta$ defined by
\begin{equation}
\delta \equiv \frac{a \sinh V_0}{2 \sinh (V_0+h)}.
\end{equation}
Setting $W=-h$ in (12.14{\it b}), we can see that
\begin{equation}
r_2 = 2 (x-\delta)^2 r_0 + \cdots,
\end{equation}
where the terms represented by the dots behave like $\rme^{-h/2}$, and hence are negligible, when $h$ is large.
In the same way, it can be shown that
\begin{equation}
r_n \simeq \frac{2^n}{n !}(x-\delta)^n r_0
\end{equation}
for any $n$, when $h$ is large.
Substituting (12.19) into (12.15) gives
\begin{equation}
R_r(x,-\infty;k) \simeq r_0\, \rme^{2 \rmi k (x-\delta)}.
\end{equation}
From (12.3), (12.20) and the second equation of (12.4), we obtain the approximation
\begin{equation}
G_{\rm S}(x,y;k)\simeq
\frac{
\left[
1+ r_0 \, \rme^{2 \rmi k (a-x-\delta)} 
\right]
\left[
1+ r_0 \, \rme^{2 \rmi k (y-\delta)}
\right]
\rme^{\rmi k (x-y)}
}
{
2 \rmi k
\left[
1 - r_0^2\, \rme^{2 \rmi k (a- 2 \delta)}
\right]
}.
\end{equation}
In fact, this approximation is equivalent to replacing the potential $V(z)$ by the effective square well potential 
\begin{equation}
V_{\rm eff}(z)=
\cases{
-h & $(\delta <z<a-\delta)$  \\
V_0 & (otherwise)
}.
\end{equation}
The result of this approximation is also plotted in figure~4 (the dotted lines).
As we can see from figures 4(b) and 4(c), equation (12.21) gives a good approximation,  when $h$ is not very small, for a wide range of $k$ except near the band edges.
This approximation becomes better for larger $h$, so that the graphs shown in figures 3(a) and 3(b) can be very accurately approximated by (12.21).

\bigskip
\noindent
{\bf Example 2.}

\nobreak
\noindent
Next, we study an example where the Schr\"odinger potential $V_{\rm S}$, rather than the Fokker-Planck potential $V$, is given. 
We consider  $V_{\rm S}(x)=V^{\rm S}_{\rm p}(x) +V^{\rm S}_\Delta(x)$ with
\numparts
\begin{equation}
V^{\rm S}_{\rm p}(x)=
\cases{
-E_0 & $(0<x<a)$ \\
C-E_0 & $(a<x<L)$ \\
},
\qquad 
V^{\rm S}_{\rm p}(x+L)=V^{\rm S}_{\rm p}(x),
\end{equation}
\begin{equation}
V^{\rm S}_\Delta(x)=
\cases{
h & $(0<x<a)$ \\
0 & (otherwise) \\
}.
\end{equation}
\endnumparts
(See figure~5.) 
The periodic part $V_{\rm p}^{\rm S}$ is a Kronig-Penny potential \cite{kittel}. 
We assume $h>0$ so that there are no bound states.
The constant $E_0$ is determined by the condition that the energy at the bottom of the lowest band be zero. It is easy to see that $E_0$ is the smallest solution of the equation
\begin{equation}
\sqrt{C-E_0} \tanh \frac{b\sqrt{C-E_0}}{2} = \sqrt{E_0} \tan \frac{a\sqrt{E_0}}{2}.
\end{equation}
The expansion of the Green function can be obtained by using the method explained in section~11. 
We define
\begin{eqnarray}
\fl
p \equiv \sqrt{E_0}, \qquad \! 
q \equiv \sqrt{C-E_0}, \qquad \! 
s \equiv \sqrt{h-E_0}, \qquad \!  
\xi \equiv \frac{q}{s} \tanh \frac{b q}{2} = \frac{p}{s} \tan \frac{a p}{2}.
\end{eqnarray}
The solution of (11.1) which remains bounded as $x \to +\infty$ is
\begin{equation}
\fl
\psi_0^+(x)=
\cases{
\cosh\, [s (x-a)] - \xi \sinh \,[s (x-a)] & $(0<x<a)$ \\
{\rm sech}\, {\textstyle \frac{b q}{2}} \cosh \left[q \left(x-a-{\textstyle \frac{b}{2}}\right)\right] & $(a < x < L)$ \\
\sec {\textstyle \frac{a p}{2}} \cos \left[p \left(x-L-{\textstyle \frac{a}{2}}\right)\right] & $(L < x < L+a)$ \\
},
\end{equation}
\begin{equation}
\psi_0^+(x+L)=\psi_0^+(x) \qquad (a<x).
\end{equation}
(We may multiply (12.26) by a positive constant factor, and this does not change the final result.) 
We do not write here the expression for $x<0$, but only remark that $\psi_0^+(x) \to +\infty$ as $x \to -\infty$. 
On account of the symmetry, $\psi_0^-$ is obtained from $\psi_0^+$ as
\begin{equation}
\psi_0^-(x)= \psi_0^+(a-x).
\end{equation}
The graphs of $V^+(x)$ and $V^-(x)$ defined by (11.2) are shown in figure~5.
%
%
%
\begin{figure}
\hspace{1cm}
\includegraphics[scale=0.75]{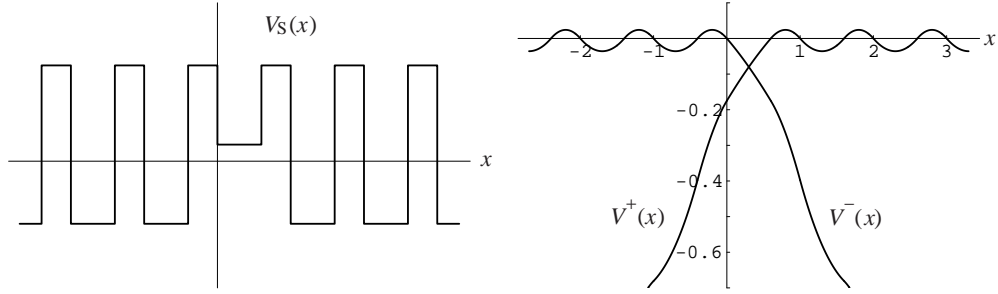}
\caption{
The Schr\"odinger potential $V_{\rm S}$ of example~2 (equations~(12.23)) and the corresponding Fokker-Planck potentials $V^\pm$.
The parameters used for these graphs are $C=1$, $L=1$, $a=0.6$ and $h=0.5$. From (12.24) we have $E_0 \simeq 0.3952$.}
\end{figure}
Note that $V^+(x) \to -\infty$ as $x \to -\infty$ and $V^-(x) \to -\infty$ as $x \to +\infty$. 

Let us calculate $g_0(x,y)$ and $g_1(x,y)$ for $0<y \leq x<a$.
We have
\begin{equation}
\fl
\psi_0^+(x)=
\cosh\, [s (x-a)] - \xi \sinh \,[s (x-a)],
\qquad
\psi_0^-(y)= \cosh s y+ \xi \sinh s y.
\end{equation}
The Wronskian can be calculated as
\begin{equation}
W[\psi_0^+, \psi_0^-] = 2 \xi s \cosh s a + (1+\xi^2) s \sinh s a.
\end{equation}
Substituting (12.29) and (12.30) into (11.9) gives $g_0(x,y)$.
To calculate $g_1$, we need to evaluate $V_0^\pm$. It is obvious that $V_0^+=V_0^-$. By definition,
\begin{equation}
\fl
\rme^{-V_0^+}=\rme^{-V_0^-}=\sqrt{M/P}, \qquad
P=\int_x^{x+L} \rme^{V_{\rm p}^+(z)}\,\rmd z, \qquad
M=\int_x^{x+L} \rme^{-V_{\rm p}^+(z)}\,\rmd z. 
\end{equation}
Since $V^+(z)=V_{\rm p}^+(z)$ for $z>a$, we can calculate, by using (12.26),
\numparts
\begin{eqnarray}
\fl
P=\int_a^{L+a}\!\!\!\frac{1}{[\psi_0^+(z)]^2}\,\rmd z
=\frac{1}{q} \sinh b q + \frac{1}{p} \sin a p,
\\
\fl
M=\int_a^{L+a} [\psi_0^+(z)]^2\,\rmd z
=\frac{b}{2}{\rm sech}^2\,\frac{b q}{2} + \frac{a}{2} \sec^2 \frac{a p}{2}
+\frac{1}{q} \tanh \frac{b q}{2} + \frac{1}{p} \tan \frac{a p}{2}.
\end{eqnarray}
\endnumparts
For $0<y\leq x<a$, we can also calculate the integrals
\begin{eqnarray}
\fl
\int_y^x \rme^{V^+(z)}\,\rmd z=
\frac{\sinh[s (x-a)]}{s \psi_0^+(x)}-\frac{\sinh[s (y-a)]}{s \psi_0^+(y)},
\qquad
\int_y^x \rme^{V^-(z)}\,\rmd z=
\frac{\sinh s x}{s \psi_0^-(x)}-\frac{\sinh s y}{s \psi_0^-(y)}.
\nonumber \\
\end{eqnarray}
Substituting the above expressions into (11.8{\it b}), we obtain 
\begin{eqnarray}
\fl
g_1(x,y)=
-\frac{1}{2}
\sqrt{
\frac{(b/2){\rm sech}^2\,\frac{b q}{2} + (a/2)\sec^2 \frac{a p}{2}
+(1/q) \tanh \frac{b q}{2} + (1/p) \tan \frac{a p}{2}}
{(1/q) \sinh b q + (1/p) \sin a p}
}
\nonumber \\
\times
\Biggl\{
\frac{\sinh[s (x-a)]}{s \psi_0^+(x)}-\frac{\sinh[s (y-a)]}{s \psi_0^+(y)}
+\frac{\sinh s x}{s \psi_0^-(x)}-\frac{\sinh s y}{s \psi_0^-(y)}
\nonumber \\
\quad \ \ 
+\frac{1}{W[\psi_0^+, \psi_0^-]}
\left(
\frac{\psi_0^+(x)}{\psi_0^-(x)}+\frac{\psi_0^-(x)}{\psi_0^+(x)}
+\frac{\psi_0^+(y)}{\psi_0^-(y)}+\frac{\psi_0^-(y)}{\psi_0^+(y)}
\right)
\Biggr\}
\frac{\psi_0^+(x)\psi_0^-(y)}{W[\psi_0^+, \psi_0^-]},
\nonumber \\
\end{eqnarray}
where $\psi_0^+(x)$, $\psi_0^-(y)$ and $W[\psi_0^+,\psi_0^-]$ are given by (12.29) and (12.30). 

The exact Green function for this potential is shown in appendix~C (equation~(C.2)). 
Comparing with this exact expression, we can check that (12.34) gives the correct first-order coefficient of the expansion (see figure~6).
%
%
%
\begin{figure}
\hspace{1cm}
\includegraphics[scale=0.75]{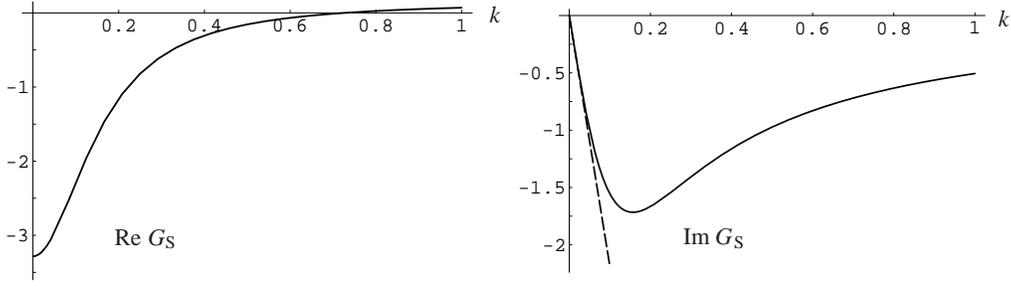}
\caption{
The real and imaginary parts of the exact $G_{\rm S}(x,y;k)$ for example~2 (equation~(C.2) of appendix~C), plotted as functions of real $k$, with $x=0.4$ and $y=0.1$. 
All other parameters 
are the same as in figure~5.
The range of $k$ shown in these graphs is  entirely included in the lowest energy band.
(The edge of the band is at $k \simeq 3.094$.)
The broken line is the line with slope $g_1\simeq -21.85$, which is the value given by (12.34). (The value of $g_0$ obtained from (11.9) is $g_0 \simeq -3.28$.) 
}
\end{figure}

\appendix
\section{Derivation of (7.8)}
We will show that 
\begin{eqnarray}
\fl
\lim_{k \to 0}
\int_{-\infty}^x
\bar Q(x,z;W;k) \,g_{\rm p}(z,\bar \omega(x,z;W;k))\,\rmd z
\nonumber \\
=\lim_{k \to 0}
\int_{-\infty}^x
\bar Q_{\rm p}(x,z;W;k) \,g_{\rm p}(z,\bar \omega_{\rm p}(x,z;W;k))\,\rmd z,
\end{eqnarray}
where $g_{\rm p}(x,W)$ is a function satisfying (4.10).
We assume that $g_{\rm p}(x,W)$ is analytic with respect to $W$ on the real axis.

Let $\bar \alpha$ and $\bar \beta$ be defined by
\begin{eqnarray}
\fl
\left(
\begin{array}{cc}
\bar \alpha(x,z;W;k) & \bar \beta(x,z;W;-k) \\
\bar \beta(x,z;W;k) & \bar \alpha(x,z;W;-k) \\
\end{array}
\right) 
\nonumber \\
\fl
\qquad \ \ 
\equiv
\left(
\begin{array}{cc}
\cosh\frac{W-V(x)}{2} &  -\sinh\frac{W-V(x)}{2} \\
-\sinh\frac{W-V(x)}{2} & \cosh\frac{W-V(x)}{2} \\
\end{array}
\right)
\left(
\begin{array}{cc}
\alpha(x,z;k) & \beta(x,z;-k) \\
\beta(x,z;k) & \alpha(x,z;-k) \\
\end{array}
\right).
\end{eqnarray}
Then, $\bar \tau$, $\bar R_r$ and $\bar R_l$ are expressed in terms of $\bar \alpha$ and $\bar \beta$ as
\begin{eqnarray}
\bar \tau(x,z;W;k)=\frac{1}{\bar \alpha(x,z;W;k)},
\qquad
\bar R_r(x,z;W;k)=\frac{\bar \beta(x,z;W;k)}{\bar \alpha(x,z;W;k)},
\nonumber \\
\bar R_l(x,z;W;k)= 
-\frac{\bar \beta(x,z;W;-k)}{\bar \alpha(x,z;W;k)}.
\end{eqnarray}
Substituting (A.3) into (7.2) and (3.6), we can write $\bar Q$ and $\bar \omega$ as
\begin{eqnarray}
\fl
\bar Q(x,z;W;k)
=
\frac{1}{[\bar \alpha(x,z;W;k) + \bar \beta(x,z;W;-k)][\bar \alpha(x,z;W;k) - \bar \beta(x,z;W;-k)]},
\nonumber \\
\bar \omega(x,z;W;k)
= V(z) 
+ \log \frac{\bar \alpha(x,z;W;k) - \bar \beta(x,z;W;-k)}{\bar \alpha(x,z;W;k) + \bar \beta(x,z;W;-k)}.
\end{eqnarray}
Let $\bar \alpha_{\rm P}$ and $\bar \beta_{\rm p}$ be the quantities obtained from $\bar \alpha$ and $\bar \beta$, respectively, by replacing $V$ with $V_{\rm p}$. Then, just like (A.3) and (A.4), we have
\begin{eqnarray}
\bar \tau^{\rm p}(x,z;W;k)=\frac{1}{\bar \alpha_{\rm P}(x,z;W;k)},
\qquad
\bar R_r^{\rm p}(x,z;W;k)=\frac{\bar \beta_{\rm P}(x,z;W;k)}{\bar \alpha_{\rm P}(x,z;W;k)},
\nonumber \\
\bar R_l^{\rm p}(x,z;W;k)= 
-\frac{\bar \beta_{\rm P}(x,z;W;-k)}{\bar \alpha_{\rm P}(x,z;W;k)}
\end{eqnarray}
and
\begin{eqnarray}
\fl
\bar Q_{\rm p}(x,z;W;k)
=
\frac{1}{[\bar \alpha_{\rm p}(x,z;W;k) + \bar \beta_{\rm p}(x,z;W;-k)][\bar \alpha_{\rm p}(x,z;W;k) - \bar \beta_{\rm p}(x,z;W;-k)]},
\nonumber \\
\bar \omega_{\rm p}(x,z;W;k)
= V_{\rm p}(z) 
+ \log \frac{\bar \alpha_{\rm p}(x,z;W;k) - \bar \beta_{\rm p}(x,z;W;-k)}{\bar \alpha_{\rm p}(x,z;W;k) + \bar \beta_{\rm p}(x,z;W;-k)}.
\end{eqnarray}

Before carrying out the calculation, let us note that the following equations hold  for an arbitrary finite number $x_0 \leq x$:
\begin{eqnarray}
\fl
\lim_{k \to 0}
\int_{-\infty}^{x} \bar Q_{\rm p}(x,z;W;k) g_{\rm p}(z, \bar \omega_{\rm p} (x,z;W;k))\, \rmd z
\nonumber \\
=
\lim_{k \to 0}
\int_{-\infty}^{x_0} \bar Q_{\rm p}(x_0,z;W;k) g_{\rm p}(z, \bar \omega_{\rm p} (x_0,z;W;k))\, \rmd z
+ \int_{x_0}^x g_{\rm p}(z,W)\,\rmd z,
\nonumber \\* 
\end{eqnarray}
\begin{eqnarray}
\fl
\lim_{k \to 0}
\int_{-\infty}^{x} \bar Q(x,z;W;k) g_{\rm p}(z, \bar \omega (x,z;W;k))\, \rmd z
\nonumber \\
=\lim_{k \to 0}
\int_{-\infty}^{x_0} \bar Q(x_0,z;W;k) g_{\rm p}(z, \bar \omega (x_0,z;W;k))\, \rmd z
+ \int_{x_0}^x g_{\rm p}(z,W)\,\rmd z.
\nonumber \\*
\end{eqnarray}
Equation (A.7) is derived from the two equations
\begin{equation}
\fl
\lim_{k \to 0}
\int_{x_0}^{x} \bar Q_{\rm p}(x,z;W;k) g_{\rm p}(z, \bar \omega_{\rm p} (x,z;W;k))\, \rmd z
=\int_{x_0}^x g_{\rm p}(z,W) \, \rmd z
\end{equation}
and
\begin{eqnarray}
\fl
\lim_{k \to 0}
\int_{-\infty}^{x_0} \bar Q_{\rm p}(x,z;W;k) g_{\rm p}(z, \bar \omega_{\rm p} (x,z;W;k))\, \rmd z
\nonumber \\
=
\lim_{k \to 0}
\int_{-\infty}^{x_0} \bar Q_{\rm p}(x_0,z;W;k) g_{\rm p}(z, \bar \omega_{\rm p} (x_0,z;W;k))\, \rmd z.
\end{eqnarray}
The proof of (A.9) is easy. Since the limit and the integral on the left-hand side are interchangeable, we can use (7.7) to obtain (A.9). 
The outline of the proof of (A.10) is as follows.
From $U(x,x_0;k) U(x_0,z;k)=U(x,z;k)$ and (A.2), we have
\begin{eqnarray}
\fl
\bar \alpha_{\rm p}(x,z;W;k) \pm \bar \beta_{\rm p}(x,z;W;-k)
=\bar \alpha_{\rm p}(x,x_0;W;k) [\alpha_{\rm p}(x_0,z;k) \pm \beta_{\rm p}(x_0,z;-k)]
\nonumber \\
+\bar \beta_{\rm p}(x,x_0;W;-k) [\beta_{\rm p}(x_0,z;k) \pm \alpha_{\rm p}(x_0,z;-k)].
\end{eqnarray}
From (2.1) and (A.2), we can easily see that
\begin{eqnarray}
\fl
\bar \alpha_{\rm p}(x,x_0;W;k)= \cosh \frac{W-V_{\rm p}(x_0)}{2} + C_1(k),
\qquad 
C_1(k)=O(k) \ \ \hbox{as} \ \  k\to 0,
\nonumber \\
\fl
\bar \beta_{\rm p}(x,x_0;W;-k)= - \sinh \frac{W-V_{\rm p}(x_0)}{2} + C_2(k),
\qquad 
C_2(k)=O(k) \ \ \hbox{as} \ \  k\to 0.
\end{eqnarray}
We substitute (A.11) and (A.12) into (A.6), and calculate the left-hand side of (A.10) by using the method described in section~5 of \cite{periodic}. Then we find that the contributions from $C_1(k)$ and $C_2(k)$ vanish in the limit $k \to 0$. 
It is easy to show that $\bar Q_{\rm p}(x,z)$ and $\bar \omega_{\rm p}(x,z)$ become $\bar Q_{\rm p}(x_0,z)$ and $\bar \omega_{\rm p}(x_0,z)$, respectively, when $C_1(k)$ and $C_2(k)$ of (A.12) are set to be zero. Hence we obtain (A.10).
Adding both sides of (A.9) and (A.10) yields (A.7).  Equation (A.8) can be proved in essentially the same way.
Using (A.7) and (A.8), equation (A.1) is reduced to
\begin{eqnarray}
\fl
\lim_{k \to 0}
\int_{-\infty}^{x_0}
\bar Q(x_0,z;W;k) \,g_{\rm p}(z,\bar \omega(x_0,z;W;k))\,\rmd z
\nonumber \\
=\lim_{k \to 0}
\int_{-\infty}^{x_0}
\bar Q_{\rm p}(x_0,z;W;k) \,g_{\rm p}(z,\bar \omega_{\rm p}(x_0,z;W;k))\,\rmd z,
\end{eqnarray}
where $x_0$ is arbitrary.
The meaning of the equivalence between (A.1) and (A.13) is easy to understand. 
Since $\lim_{k \to 0} \bar Q = \lim_{k \to 0} \bar Q_{\rm p}$ and $\lim_{k \to 0} \bar \omega = \lim_{k \to 0} \bar \omega_{\rm p}$, the difference between the left-hand and right-hand sides of (A.1) in the limit $k \to 0$ is determined only by the behavior at $z \to -\infty$. So, in (A.1), we can arbitrarily change the value of $x$ as long as it remains finite. Also note that equations (A.7) and (A.8), respectively, correspond to
\begin{eqnarray}
(\mathcal{A}_{\rm p}^{-1} g_{\rm p})(x,W)=(\mathcal{A}_{\rm p}^{-1} g_{\rm p})(x_0,W)
+ \int_{x_0}^x g_{\rm p}(z,W)\, \rmd z,
\nonumber \\
(\mathcal{A}^{-1} g_{\rm p})(x,W)=(\mathcal{A}^{-1} g_{\rm p})(x_0,W)
+ \int_{x_0}^x g_{\rm p}(z,W)\, \rmd z
\end{eqnarray} 
(see equation (5.34) of \cite{periodic}). 

Since (A.1) and (A.13) are equivalent, we will deal with (A.13) instead of (A.1).
By taking $-x_0$ to be sufficiently large, we can make $\vert V_\Delta(z) \vert$ to be arbitrarily small for $z \leq x_0$.

Let us review how to calculate the right-hand side of (A.13). (See \cite{periodic} for details.)
We split the integral into unit periods as
\begin{eqnarray}
\int_{-\infty}^{x_0}
\bar Q_{\rm p}(x_0,z;W;k) \,g_{\rm p}(z,\bar \omega_{\rm p}(x_0,z;W;k))\,\rmd z
=\sum_{n=0}^\infty A_n(k),
\\
A_n(k)
\equiv
\int_{x_0- (n+1) L}^{x_0 - n L}
\bar Q_{\rm p}(x_0,z;W;k) \,g_{\rm p}(z,\bar \omega_{\rm p}(x_0,z;W;k))\,\rmd z.
\end{eqnarray}
Let $\lambda$ and $\lambda^{-1}$ be the eigenvalues of the matrix
\begin{equation}
\left(
\begin{array}{cc}
\alpha_{\rm p}(x,x - L;k) & \beta_{\rm p}(x,x - L;-k) \\
\beta_{\rm p}(x,x - L;k) & \alpha_{\rm p}(x,x - L;-k) \\
\end{array}
\right)
\end{equation}
such that $\vert \lambda \vert \geq 1$ for ${\rm Im}\,k \geq 0$. Namely,
\begin{equation}
\fl
\lambda \equiv Y- \rmi \sqrt{1- Y^2},
\qquad
Y \equiv \frac{1}{2}\left[\alpha_{\rm p}(x,x-L;k)+\alpha_{\rm p}(x,x-L;-k)\right].
\end{equation}
We define
\begin{equation}
\gamma \equiv \lambda^{-2}.
\end{equation} 
It turns out that $A_n(k)$ depends on $n$ only through $\gamma^n$. Moreover, it also turns out that $A_n(k)$ has the form
\begin{equation}
A_n(k)=k \gamma^n C(\gamma^n, k),
\end{equation}
where $C(\gamma^n, k)$ is an analytic function of $k$ and $\gamma^n$ on the real axis.
We have $\vert \gamma \vert <1$ for ${\rm Im}\,k>0$ and $\vert \gamma \vert =1$ for ${\rm Im}\,k =0$. 
The behavior of $\gamma$ for small $k$ is
\begin{equation}
\gamma = 1 + 2 \rmi L_0 k + O(k^2) \qquad \hbox{as} \quad k \to 0.
\end{equation}
Since the approach of $\gamma^n$ to 1 is not uniform in $n$, we cannot change the order of the limit and the sum as 
\begin{equation}
\lim_{k \to 0} \sum_{n=0}^\infty k \gamma^n C(\gamma^n, k)
=
\sum_{n=0}^\infty \lim_{k \to 0} k \gamma^n C(\gamma^n, k)=0.
\end{equation}
However, if we treat $k$ and $\gamma^n$ separately, we can let $k\to 0$ inside the sum as
\begin{equation}
\lim_{k \to 0} k \sum_{n=0}^\infty \gamma^n C(\gamma^n, k)
=
\lim_{k \to 0} k \sum_{n=0}^\infty \gamma^n C(\gamma^n, 0).
\end{equation}
As a result, we have
\begin{equation}
\lim_{k \to 0} \, k \sum_{n=0}^\infty \gamma^n C(\gamma^n, k)
=
\frac{\rmi}{2 L_0} \int_0^1 C(x,0) \, \rmd x.
\end{equation}
This equation can be intuitively understood as follows.
From (A.21), we can see that $\gamma^n$ behaves like $\exp(2 \rmi L_0 k n)$ when $k$ is small. 
Therefore,
\begin{eqnarray}
\fl
\lim_{k \to 0}\,k \sum_{n=0}^\infty \gamma^n C(\gamma^n, k)
=
\lim_{k=0}\,k \int_0^\infty  \rme^{2 \rmi L_0 k \nu}
C(\rme^{2 \rmi L_0 k \nu}, k) \, \rmd \nu
= \frac{\rmi}{2 L_0}\lim_{k \to 0}\int_0^1 C(x,k)\, \rmd x.
\nonumber \\*
\end{eqnarray} 
In (A.25), we changed the sum over discrete $n$ to an integral over continuous $\nu$, and then changed the variable of integration to $x \equiv \rme^{2 \rmi L_0 k \nu}$.
Here, let us comment about the meaning of $\lim_{k\to 0}$.  The limit $k\to 0$ can be taken in any way in the closed upper half plane (${\rm Im}\,k\geq 0$), but we need to be careful when the limit is taken along the real axis (${\rm Im}\,k =0$). 
Recall that we have defined the Green function for ${\rm Im}\,k=0$ as $G_{\rm S}(x,y;k)\equiv \lim_{\epsilon \downarrow 0}G_{\rm S}(x,y;k+\rm i\epsilon)$. 
So when we write an expression like $\lim_{k \to 0} a(k)$, it is to be understood as $\lim_{k \to 0} \lim_{\epsilon \downarrow 0} a(k + \rmi \epsilon)$ for real $k$. Therefore, in (A.25), $x \to 0 $ as $\nu \to \infty$ inside the limit sign. 
The limit and the integral in the last expression of (A.25) can be interchanged, since $\vert C(x,k) \vert$ is uniformly bounded when $\vert k \vert$ is sufficiently small. Hence we have (A.24). 

The left-hand side of (A.13) is calculated in almost the same way. 
The only difference is that the quantity corresponding to $A_n(k)$ contains a part which depends explicitly on $n$, in addition to the part which depends on $n$ through $\gamma^n$. If we write
\begin{eqnarray}
\int_{-\infty}^{x_0}
\bar Q(x_0,z;W;k) \,g_{\rm p}(z,\bar \omega(x_0,z;W;k))\,\rmd z
\equiv \sum_{n=0}^\infty A'_n(k),
\end{eqnarray}
then $A'_n(k)$ has the form
\begin{equation}
A'_n(k)=k \gamma^n C(\gamma^n, k) + D_n(\gamma^n,k).
\end{equation}
As we will see, $D_n$ decreases like $V_\Delta(x_0 - n L)$ as $n \to \infty$. If $V_\Delta \in F^{(-)}_0$, 
 there exists a $k$-independent (and $\gamma^n$-independent) sequence $\{E_n\}$ such that $\vert D_n(\gamma^n,k) \vert < E_n$ and $\sum_{n=0}^\infty E_n < \infty$. Then, we have
\begin{equation}
\lim_{k\to 0} \sum_{n=0}^\infty D_n(\gamma^n,k) = \sum_{n=0}^\infty D_n(1,0).
\end{equation}
We will show that the right-hand side of (A.28) is zero. 

Now let us explicitly calculate the difference between the right-hand and left-hand sides of (A.13). Since $\vert V_\Delta \vert$ can be made as small as we like by taking large $\vert x_0 \vert$, we consider only the terms of first order in $V_\Delta$. We can write
\begin{eqnarray}
\fl
\int_{-\infty}^{x_0} \bar Q(x_0,z;W;k) g_{\rm p}(z, \bar \omega (x_0,z;W;k))\, \rmd z
\nonumber \\
\quad 
- \int_{-\infty}^{x_0} \bar Q_{\rm p}(x_0,z;W;k) g_{\rm p}(z, \bar \omega_{\rm p} (x_0,z;W;k))\, \rmd z
\nonumber \\
= \int_{-\infty}^{x_0} \bar Q_\Delta (x_0,z;W;k) g_{\rm p}(z, \bar \omega_{\rm p} (x_0,z;W;k))\, \rmd z
\nonumber \\
\quad
+ \int_{-\infty}^{x_0} \bar Q_{\rm p}(x_0,z;W;k) 
\bar \omega_\Delta (x_0,z;W;k) g'_{\rm p}(z, \bar \omega_{\rm p} (x_0,z;W;k))\, \rmd z,
\end{eqnarray}
where
\begin{equation}
\bar Q_\Delta \equiv \bar Q -\bar Q_{\rm p}, 
\qquad
\bar \omega_\Delta \equiv \bar \omega - \bar \omega_{\rm p},
\qquad
g'_{\rm p}(z,W) \equiv \frac{\partial}{\partial W} g_{\rm p}(z,W).
\end{equation}
We define
\begin{eqnarray}
\zeta_{\rm p}(z) \equiv \bar \alpha_{\rm p}(x_0, z; W; k) + \bar \beta_{\rm p}(x_0, z; W; -k),
\nonumber \\
\eta_{\rm p}(z) \equiv \bar \alpha_{\rm p}(x_0, z; W; k) - \bar \beta_{\rm p}(x_0, z; W; -k)
\end{eqnarray}
and
\begin{eqnarray}
\zeta(z) \equiv \bar \alpha(x_0, z; W; k) + \bar \beta(x_0, z; W; -k),
\nonumber \\
\eta(z) \equiv \bar \alpha(x_0, z; W; k) - \bar \beta(x_0, z; W; -k).
\end{eqnarray}
Then, as can be seen from (A.4) and (A.6),
\begin{eqnarray}
\fl
\bar Q_{\rm P}(x_0,z;W;k)=\frac{1}{\zeta_{\rm P}(z) \eta_{\rm P(}z)},
\qquad
& \bar \omega_{\rm P}(x_0,z;W;k)=V_{\rm P}(z) + \log \frac{\eta_{\rm P}(z)}{\zeta_{\rm P(}z)},
\\ 
\fl
\bar Q(x_0,z;W;k)=\frac{1}{\zeta(z) \eta(z)},
\qquad
& \bar \omega(x_0,z;W;k)=V(z) + \log \frac{\eta(z)}{\zeta(z)}.
\end{eqnarray}
We can express $\bar Q_\Delta$ and $\bar \omega_\Delta$ as
\begin{equation}
\bar Q_\Delta = -\frac{1}{\zeta_{\rm p} \eta_{\rm p}}
\left(\frac{\zeta_\Delta}{\zeta_{\rm p}}
+ \frac{\eta_\Delta}{\eta_{\rm p}} \right),
\qquad
\bar \omega_\Delta = V_\Delta - \frac{\zeta_\Delta}{\zeta_{\rm p}} 
+ \frac{\eta_\Delta}{\eta_{\rm p}},
\end{equation}
where
\begin{equation}
\zeta_\Delta(z) \equiv \zeta(z) - \zeta_{\rm p}(z), 
\qquad
\eta_\Delta(z) \equiv \eta(z) - \eta_{\rm p}(z).
\end{equation}
Substituting  (A.35) into (A.29) gives
\begin{eqnarray}
\fl
\int_{-\infty}^{x_0} \bar Q(x_0,z;W;k) g_{\rm p}(z, \bar \omega (x_0,z;W;k))\, \rmd z
\nonumber \\
\quad 
- \int_{-\infty}^{x_0} \bar Q_{\rm p}(x_0,z;W;k) g_{\rm p}(z, \bar \omega_{\rm p} (x_0,z;W;k))\, \rmd z
\nonumber \\
= \int_{-\infty}^{x_0} \frac{-\zeta_\Delta (z)}{\zeta_{\rm p}^2(z)\eta_{\rm p}(z)}
\left[
g_{\rm p}(z,\bar \omega_{\rm p}) + g'_{\rm p}(z,\bar \omega_{\rm p})
\right] \rmd z
\nonumber \\
\quad
+
\int_{-\infty}^{x_0} \frac{-\eta_\Delta (z)}{\zeta_{\rm p}(z)\eta_{\rm p}^2(z)}
\left[
g_{\rm p}(z,\bar \omega_{\rm p}) - g'_{\rm p}(z,\bar \omega_{\rm p})
\right] \rmd z
\nonumber \\
\quad
+ \int_{-\infty}^{x_0} \frac{V_\Delta (z)}{\zeta_{\rm p}(z)\eta_{\rm p}(z)}\,
g'_{\rm p}(z,\bar \omega_{\rm p})\, \rmd z.
\end{eqnarray}
As in (A.15) and (A.16), we calculate each integral for $x_0- n L \geq z \geq x_0- (n+1)L$, and then take the sum over $n$. So we need the expressions of $\zeta_{\rm p}(z)$, $\eta_{\rm p}(z)$, $\zeta_\Delta(z)$ and $\eta_\Delta(z)$ for $x_0- n L \geq z \geq x_0- (n+1)L$.
It is shown in \cite{periodic} that
\begin{eqnarray}
\fl
\zeta_{\rm p}(z)= \rme^{u(z)} (\cosh w \,\lambda^n - \sinh w\, \lambda^{-n})
- \rmi k \rme^{v(z)} (\cosh w \,\lambda^n + \sinh w\, \lambda^{-n}) \,{}_{\rm p}[\hbox{$-$}]_z^{x_0-n L}
\nonumber \\
+ \rme^u C(\lambda) k + o(k),
\nonumber \\
\fl
\eta_{\rm p}(z)= \rme^{-u(z)} (\cosh w \,\lambda^n + \sinh w\, \lambda^{-n})
- \rmi k \rme^{-v(z)} (\cosh w \,\lambda^n - \sinh w\, \lambda^{-n}) \,{}_{\rm p}[\hbox{$+$}]_z^{x_0-n L}
\nonumber \\
+ \rme^{-u(z)} C(\lambda) k + o(k),
\end{eqnarray}
where $w$, $u$ and $v$ are defined by
\begin{equation}
w \equiv \frac{W-V_0}{2}, \qquad
v(z) \equiv \frac{V_{\rm p}(z) + V_0}{2}, \qquad
u(z) \equiv \frac{V_{\rm p}(z) - V_0}{2}.
\end{equation}
In (A.38), and in equations (A.40), we let $C(\lambda)$ stand for a $z$-independent function of $\lambda$ which is not necessarily the same everywhere.
The expressions for $\zeta_\Delta(z)$ and $\eta_\Delta(z)$ can be obtained, after some calculation,  as
\begin{eqnarray}
\fl
\zeta_\Delta(z)=
\rme^{u(z)}
\sum_{m=0}^{n-1} B_m (\sinh w \,\lambda^{n-2m-1} - \cosh w\, \lambda^{-n+2m+1})
\nonumber \\
- \rmi k \rme^{v(z)}\,{}_{\rm p}[\hbox{$-$}]_z^{x_0-n L}
\sum_{m=0}^{n-1} B_m (\sinh w \,\lambda^{n-2m-1} + \cosh w\, \lambda^{-n+2m+1})
\nonumber \\
-\frac{1}{2}V_\Delta(x_0) 
\,\rme^{u(z)} (\sinh w \,\lambda^n - \cosh w\, \lambda^{-n})
\nonumber \\
+ \frac{\rmi k }{2}V_\Delta(x_0)\, \rme^{v(z)} \,{}_{\rm p}[\hbox{$-$}]_z^{x_0-n L}\, (\sinh w \,\lambda^n + \cosh w\, \lambda^{-n}) 
\nonumber \\
+ \frac{1}{2} \left[V_\Delta(z) - V_\Delta(x_0-n L)\right]
\,\rme^{u(z)} (\cosh w \,\lambda^n - \sinh w\, \lambda^{-n})
\nonumber \\
+ \rme^{u(z)} C(\lambda) k + o(k),
\nonumber \\
\fl
\eta_\Delta(z)=
\rme^{- u(z)}
\sum_{m=0}^{n-1} B_m (\sinh w \,\lambda^{n-2m-1} + \cosh w\, \lambda^{-n+2m+1})
\nonumber \\
- \rmi k \rme^{-v(z)}\,{}_{\rm p}[\hbox{$+$}]_z^{x_0-n L}
\sum_{m=0}^{n-1} B_m (\sinh w \,\lambda^{n-2m-1} - \cosh w\, \lambda^{-n+2m+1})
\nonumber \\
-\frac{1}{2}V_\Delta(x_0) 
\,\rme^{-u(z)} (\sinh w \,\lambda^n + \cosh w\, \lambda^{-n})
\nonumber \\
+ \frac{\rmi k }{2}V_\Delta(x_0)\, \rme^{-v(z)} \,{}_{\rm p}[\hbox{$+$}]_z^{x_0-n L}\, (\sinh w \,\lambda^n - \cosh w\, \lambda^{-n}) 
\nonumber \\
- \frac{1}{2} \left[V_\Delta(z) - V_\Delta(x_0-n L)\right]
\,\rme^{-u(z)} (\cosh w \,\lambda^n + \sinh w\, \lambda^{-n})
\nonumber \\
+ \rme^{-u(z)}C(\lambda) k + o(k),
\end{eqnarray}
where
\begin{equation}
B_m \equiv \frac{1}{2} \left[V_\Delta(x_0 - mL) - V_\Delta(x_0 - (m+1) L) \right].
\end{equation}

Let us consider the first term on the right-hand side of (A.37).
We divide the integral into unit periods and write
\begin{eqnarray}
\fl
\int_{-\infty}^{x_0} \frac{-\zeta_\Delta (z)}{\zeta_{\rm p}^2(z)\eta_{\rm p}(z)}\,
g_{\rm p}(z,\bar \omega_{\rm p})\, \rmd z
=
\sum_{n=0}^\infty
\int_{x_0- (n+1) L}^{x_0 - n L}\frac{-\zeta_\Delta (z)}{\zeta_{\rm p}^2(z)\eta_{\rm p}(z)}\,
g_{\rm p}(z,\bar \omega_{\rm p})\, \rmd z
\nonumber \\
=
\sum_{n=0}^\infty
\int_{x_0- (n+1) L}^{x_0 - n L} a(z,k) \, b(z,k)\, d(z,k)\, \rmd z,
\end{eqnarray}
where we have defined
\begin{equation}
\fl
a(z,k) \equiv \frac{\rme^{u(z)} \lambda^n \cosh w}{\zeta_{\rm p}^2(z) \eta_{\rm p}(z)},
\qquad
b(z,k) \equiv g_{\rm p}(z, \bar \omega_{\rm p}),
\qquad
d(z,k) \equiv \frac{- \zeta_\Delta(z)}{\rme^{u(z)} \lambda^n \cosh w}.
\end{equation}
From (A.38) and (A.40), we obtain the expressions for $a(z,k)$, $b(z,k)$ and $d(z,k)$ as
\begin{eqnarray}
a(z,k)=a_0(\gamma^n) + k a_1(z,\gamma^n) + o(k),
\nonumber \\
b(z,k)=b_0(z) + k b_1(z, \gamma^n) + o(k),
\nonumber \\
d(z,k)=d_0(z,\gamma) + k d_1(z,\gamma) + o(k),
\end{eqnarray}
\begin{eqnarray}
\fl
a_0(\gamma^n)=\frac{\gamma^n}{(\cosh w)^2 (1-c_0 \gamma^n)(1 + c_0 \gamma^n)^2},
\nonumber \\
\fl
a_1(z,\gamma^n)= \rmi a_0(\gamma^n)
\left(
\rme^{V_0} \frac{1 - c_0 \gamma^n}{1 + c_0 \gamma^n}\,
{}_{\rm p}[\hbox{$-$}]_z^{x_0-n L}
+ 
\rme^{-V_0} \frac{1 + c_0 \gamma^n}{1 - c_0 \gamma^n}\,
{}_{\rm p}[\hbox{$+$}]_z^{x_0-n L}
\right) + C(\gamma^n),
\end{eqnarray}
\begin{eqnarray}
\fl
b_0(z)= g_{\rm p}(z,W),
\nonumber \\
\fl
b_1(z,\gamma^n)
=\rmi g'_{\rm p}(z,W)
\left(
\rme^{V_0} \frac{1 - c_0 \gamma^n}{1 + c_0 \gamma^n}\,
{}_{\rm p}[\hbox{$-$}]_z^{x_0-n L}
-
\rme^{-V_0} \frac{1 + c_0 \gamma^n}{1 - c_0 \gamma^n}\,
{}_{\rm p}[\hbox{$+$}]_z^{x_0-n L}
\right) + C(\gamma^n), 
\nonumber \\*
\end{eqnarray}
\begin{eqnarray}
\fl
d_0(z,\gamma)
= 
\sum_{m=0}^{n-1} B_m \left(c_0 \gamma^m + \gamma^{n-m}\right)
-\frac{1}{2}V_\Delta(x_0) \left(c_0 + \gamma^n \right)
\nonumber \\
\quad
-\frac{1}{2} \left[ V_\Delta(z) - V_\Delta(x_0 - n L)\right]\left(1 + c_0 \gamma^n \right),
\nonumber \\
\fl
d_1(z,\gamma)=
- \rmi \rme^{V_0} {}_{\rm p}[\hbox{$-$}]_z^{x_0-n L}
\sum_{m=0}^{n-1} B_m \left(c_0 \gamma^m - \gamma^{n-m}\right)
\nonumber \\
\quad 
+\frac{\rmi}{2}V_\Delta(x_0) \, \rme^{V_0} {}_{\rm p}[\hbox{$-$}]_z^{x_0-n L}
\left(c_0 - \gamma^n \right) + C(\gamma),
\end{eqnarray}
where 
\begin{equation}
c_0 \equiv -\tanh w.
\end{equation}
In the above expressions, $C(\gamma^n)$ and $C(\gamma)$ stand for the terms independent of $z$, as in equations (A.38) and (A.40).

As explained before, we can safely ignore the $o(k)$ terms in (A.44). For any fixed finite number $l$, we can expand $\gamma^l$ in terms of $k$ and neglect the higher-order terms, but we cannot do so for $\gamma^n$. We must leave $\gamma^n$ as it is.
We can see that $a_0$, $a_1$ and $b_1$ depend on $\gamma$ only through $\gamma^n$. However, this is not the case for $d_0$ and $d_1$. As shown in (A.47), they have the terms including $\sum_{m=0}^{n-1} B_m \gamma^m$ and $\sum_{m=0}^{n-1} B_m \gamma^{n-m}$. We need to know how to deal with these terms. The conclusion is that $\gamma^m$ and $\gamma^{-m}$ in these terms can be replaced by 1. Let us explain why this is so.
Substituting (A.44) with (A.47) into the last expression of (A.42), we have the terms involving $\gamma^m$ and $\gamma^{-m}$ as
\begin{equation}
\sum_{n=0}^\infty S_n(k) \sum_{m=0}^{n-1} B_m (c_0 \gamma^m + \gamma^{n-m}),
\end{equation}
where
\begin{equation}
\fl
S_n(k)=\int_{x_0 - (n+1) L}^{x_0 - n L} a(z,k) b(z,k)
\left( 1 -  \rmi k \rme^{-V_0} {}_{\rm p}[\hbox{$-$}]_z^{x_0-n L}\right) \, \rmd z.
\end{equation}
On the right-hand side of (A.50), the terms of order $k^0$ vanish since $\int a_0(\gamma^n) b_0(z)\, \rmd z=0$ on account of (4.10). So, $S_n(k)$ has the form
\begin{equation}
S_n(k)=k \gamma^n h(\gamma^n),
\end{equation}
where $h$ is an analytic function which can be expanded as $h(x)=h_0 + h_1 x + h_2 x^2 +\cdots$. 
(Note that $a(z,k)$ has a factor $\gamma^n$, as shown in (A.45). Hence comes the factor $\gamma^n$ in front of $h(\gamma^n)$ in (A.51).)
We can show that
\begin{eqnarray}
\lim_{k \to 0} k \sum_{n=0}^\infty \gamma^n h(\gamma^n) \sum_{m=0}^{n-1} B_m \gamma^m 
=
\lim_{k \to 0} k \sum_{n=0}^\infty \gamma^n h(\gamma^n) \sum_{m=0}^{n-1} B_m,
\\
\lim_{k \to 0} k \sum_{n=0}^\infty \gamma^n h(\gamma^n) \sum_{m=0}^{n-1} B_m \gamma^{n-m} 
=
\lim_{k \to 0} k \sum_{n=0}^\infty \gamma^n h(\gamma^n) \sum_{m=0}^{n-1} B_m \gamma^n.
\end{eqnarray}
Equation (A.52) is proved as follows. 
Since we can express $h(\gamma^n)$ as a power series in terms of $\gamma^n$, it is sufficient to consider the case $h(\gamma^n)=\gamma^{l n}$ with an integer $l \geq 0$. 
We need to show that
\begin{equation}
\lim_{k \to 0} k \sum_{n=0}^\infty \gamma^{n (l+1)} \sum_{m=0}^{n-1} B_m \gamma^m
=
\lim_{k \to 0} k \sum_{n=0}^\infty \gamma^{n (l+1)} \sum_{m=0}^{n-1} B_m.
\end{equation}
We rewrite the left-hand side of (A.54) as
\begin{equation}
\lim_{k \to 0} k \sum_{n=0}^\infty \gamma^{n (l+1)} \sum_{m=0}^{n-1} B_m \gamma^m
=
\lim_{k \to 0} k \sum_{m=0}^\infty B_m \gamma^m \sum_{n=m+1}^\infty \gamma^{n (l+1)}.
\end{equation}
Since $\vert \gamma \vert < 1$ for ${\rm Im}\, k>0$, we can calculate the sum over $n$ on the right-hand side. 
(As mentioned before, when the limit $k \to 0$ of a function $a(k)$ is taken along the real axis, it should be understood as $\lim_{k\to 0} a(k)= \lim_{k \to 0} \lim_{\epsilon \downarrow 0} a(k + \rmi \epsilon)$. So we may assume $\vert \gamma \vert <1$ before taking the limit, even when we are considering real $k$.) We obtain
\begin{equation}
\lim_{k \to 0} k \sum_{m=0}^\infty B_m \gamma^m \sum_{n=m+1}^\infty \gamma^{n (l+1)}
=\lim_{k \to 0} k \frac{\gamma^{l+1}}{1-\gamma^{l+1}}
\sum_{m=0}^\infty B_m \gamma^{m (l+2)}.
\end{equation}
From (A.21) we have
\begin{equation}
\lim_{k \to 0} k \frac{\gamma^{l+1}}{1-\gamma^{l+1}}= \frac{\rmi}{2 (l+1) L_0}.
\end{equation}
Since $\vert \gamma^{m (l+2)} \vert \leq 1$ for ${\rm Im}\,k \geq 0$, 
and since $\sum_{m=0}^\infty \vert B_m \vert < \infty$, we have
\begin{equation}
\lim_{k\to 0} \sum_{m=0}^\infty B_m \gamma^{m (l+1)}=\sum_{m=0}^\infty B_m.
\end{equation}
Therefore, 
\begin{equation}
\lim_{k \to 0} k \sum_{n=0}^\infty \gamma^{n (l+1)} \sum_{m=0}^{n-1} B_m \gamma^m
= \frac{\rmi}{2 (l+1) L_0} \sum_{m=0}^\infty B_m.
\end{equation}
It is obvious that the right-hand side of (A.59) is equal to the right-hand side of (A.54). 
Hence, we obtain (A.54).

It is also possible to derive (A.52) by using the same argument as in (A.25).
Let us define
\begin{equation}
B(\mu) \equiv B_m \qquad \hbox{for} \quad  m\leq \mu < m+1,
\end{equation}
\begin{equation}
F(k \nu) \equiv k \int_{\infty}^\nu \rme^{2 \rmi L_0 k \nu} h(\rme^{2 \rmi L_0 k \nu})\,\rmd \nu 
= \frac{-\rmi}{2 L_0} \int_0^{\exp(2 \rmi L_0 k \nu)}\!  h(x) \, \rmd x.
\end{equation}
Then,
\begin{eqnarray}
\fl
\lim_{k \to 0} k \sum_{n=0}^\infty \gamma^n h(\gamma^n) \sum_{m=0}^{n-1} B_m \gamma^m 
=
\lim_{k \to 0} k 
\int_0^\infty \rme^{2 \rmi L_0 k \nu} h(\rme^{2 \rmi L_0 k \nu})
\left[
\int_0^\nu B(\mu) \rme^{2 \rmi L_0 k \mu}\, \rmd \mu
\right] \rmd \nu
\nonumber \\
=
\lim_{k \to 0}  
\int_0^\infty \frac{\partial F(k \nu)}{\partial \nu}
\left[
\int_0^\nu B(\mu) \rme^{2 \rmi L_0 k \mu}\, \rmd \mu
\right] \rmd \nu
\nonumber \\
=
\lim_{k \to 0} 
\left\{
F(k \nu) \int_0^\nu B(\mu) \rme^{2 \rmi L_0 k \mu}\, \rmd \mu \,\Biggl\vert_{\nu=0}^{\nu=\infty}
\right\}
- 
\lim_{k \to 0} 
\int_0^\infty F(k \nu) B(\nu) \rme^{2 \rmi L_0 k \nu}\, \rmd \nu.
\nonumber \\*
\end{eqnarray}
It can be seen from (A.61) that
\begin{equation}
F(\infty)=0, \qquad F(0)=\frac{-\rmi}{2 L_0} \int_0^1 h(x) \,\rmd x.
\end{equation}
 (When $k$ is real, $k$ is replaced by $k + \rmi \epsilon$ with positive infinitesimal $\epsilon$.) So the first term on the last line of (A.62) vanishes. 
 In the second term, the limit and the integral can be interchanged since $\vert F(k \nu)  \rme^{2 \rmi L_0 k \nu}\vert$ is uniformly bounded 
 and $\int_{-\infty}^\infty \vert B(\nu) \vert \,\rmd \nu < \infty$. Hence,
\begin{eqnarray}
\fl
\lim_{k \to 0} k \sum_{n=0}^\infty \gamma^n h(\gamma^n) \sum_{m=0}^{n-1} B_m \gamma^m 
=
-\int_0^\infty \lim_{k \to 0} F(k \nu) B(\nu) \rme^{2 \rmi L_0 k \nu}\, \rmd \nu
\nonumber \\
=
-F(0) \int_0^\infty B(\nu)\,\rmd \nu 
=
\frac{\rmi}{2 L_0} \int_0^1 h(x) \,\rmd x
 \sum_{m=0}^\infty B_m.
\end{eqnarray}
This is the same result as (A.59). Since the last expression of (A.64) is obviously equal to the right-hand side of (A.52), we can see that (A.52) holds.
Equation (A.53) can be proved in the same way. We have
\begin{eqnarray}
\fl
\lim_{k \to 0} k \sum_{n=0}^\infty \gamma^{n (l+1)} \sum_{m=0}^{n-1} B_m \gamma^{n-m}
=
\lim_{k \to 0} k \frac{\gamma^{l+2}}{1-\gamma^{l+2}}
\sum_{m=0}^\infty B_m \gamma^{m (l+1)}
= \frac{\rmi}{2 (l+2) L_0} \sum_{m=0}^\infty B_m
\nonumber \\*
\end{eqnarray}
and
\begin{eqnarray}
\fl
\lim_{k \to 0} k \sum_{n=0}^\infty \gamma^n h(\gamma^n) \sum_{m=0}^{n-1} B_m \gamma^{n-m} 
=
\frac{\rmi}{2 L_0} \int_0^1 x h(x) \,\rmd x
 \sum_{m=0}^\infty B_m
 \nonumber \\
 =
 \lim_{k \to 0} k \sum_{n=0}^\infty \gamma^n h(\gamma^n) \sum_{m=0}^{n-1} B_m \gamma^n.
\end{eqnarray}
Thus, we obtain
\begin{equation}
\fl
\lim_{k \to 0}\sum_{n=0}^\infty S_n(k) \sum_{m=0}^{n-1} B_m (c_0 \gamma^m + \gamma^{n-m})
=
\lim_{k \to 0}\sum_{n=0}^\infty S_n(k) \sum_{m=0}^{n-1} B_m (c_0 + \gamma^n).
\end{equation}
Namely, $\gamma^m$ and $\gamma^{-m}$ in (A.49) can be replaced by 1.

Now we know that we can replace every $\gamma$ by $1$, as long as we keep $\gamma^n$ aside.
Let $\tilde d(z,k)$ be the quantity obtained from $d(z,k)$ by this replacement. Then,
\begin{eqnarray}
\fl
\lim_{k \to 0}
\sum_{n=0}^\infty
\int_{x_0- (n+1) L}^{x_0 - n L} a(z,k) \,b(z,k)\, d(z,k)\, \rmd z
=
\lim_{k \to 0}
\sum_{n=0}^\infty
\int_{x_0- (n+1) L}^{x_0 - n L} a(z,k) \,b(z,k)\, \tilde d(z,k)\, \rmd z.
\nonumber \\*
\end{eqnarray}
With $\gamma^{\pm m} \to 1$, we have
\begin{eqnarray}
\sum_{m=0}^{n-1} B_m \gamma^m \to
\sum_{m=0}^{n-1} B_m 
=\frac{1}{2}[V_\Delta(x_0) - V_\Delta(x_0 - n L)] ,
\nonumber \\
\sum_{m=0}^{n-1} B_m \gamma^{n-m} \to
\sum_{m=0}^{n-1} B_m  \gamma^n
=\frac{1}{2}[V_\Delta(x_0) - V_\Delta(x_0 - n L)] \gamma^n,
\end{eqnarray}
and so equations (A.47) become
\begin{equation}
d_0(z,\gamma) \to \tilde d_\Delta(z,\gamma^n) + \tilde d_0(\gamma^n),
\qquad
d_1(z,\gamma) \to \tilde d_1(z,\gamma^n) + C(\gamma^n),
\end{equation}
where
\begin{eqnarray}
\fl
\tilde d_\Delta(z, \gamma^n) \equiv -\frac{1}{2} V_\Delta(z) (1 + c_0 \gamma^n),
\qquad
\tilde d_0(\gamma^n) \equiv  \frac{1}{2}V_\Delta(x_0- n L)  (1 - c_0)(1 - \gamma^n),
\nonumber \\
\tilde d_1(z, \gamma^n)
\equiv \frac{\rmi}{2}V_\Delta(x_0- n L) \, \rme^{V_0} {}_{\rm p}[\hbox{$-$}]_z^{x_0-n L}
\left(c_0 - \gamma^n \right),
\end{eqnarray}
and $C(\gamma^n)$ is an $z$-independent term.
With (A.71), we can write $\tilde d(z,k)$ as
\begin{equation}
\tilde d(z,k)=\tilde d_\Delta(z, \gamma^n) + \tilde d_0(\gamma^n) + k \tilde d_1(z,\gamma^n) +  k C(\gamma^n).
\end{equation}

We substitute (A.44) and (A.72) (with  (A.45), (A.46) and (A.71)) into the right-hand side of (A.68), and calculate the integral over $z$. 
Since $g_{\rm p}$ satisfies (4.10), we have 
\begin{eqnarray}
\fl
\int_{x_0-(n+1) L}^{x_0 - n L} a_0(\gamma^n) b_0(z) \tilde d_0(\gamma^n) \,\rmd z
=a_0(\gamma^n) \tilde d_0(\gamma^n) \int_{x_0-(n+1) L}^{x_0 - n L} g_{\rm p}(z,W) \,\rmd z=0,
\nonumber \\
\fl
\int_{x_0-(n+1) L}^{x_0 - n L} a_0(\gamma^n) b_0(z) C(\gamma^n) \,\rmd z
=a_0(\gamma^n) C(\gamma^n) \int_{x_0-(n+1) L}^{x_0 - n L} g_{\rm p}(z,W) \,\rmd z=0.
\end{eqnarray}
Therefore,
\begin{equation}
\int_{x_0- (n+1) L}^{x_0 - n L} a(z,k) \,b(z,k)\, \tilde d(z,k)\, \rmd z
= F_n(\gamma^n) + k G_n(\gamma^n,k),
\end{equation}
where we have defined
\begin{equation}
F_n(\gamma^n) \equiv a_0(\gamma^n) \int_{x_0- (n+1) L}^{x_0 - n L} b_0(z)\, \tilde d_\Delta(z,\gamma^n)\, \rmd z,
\end{equation}
\begin{eqnarray}
\fl
G_n(\gamma^n,k) \equiv \tilde d_0(\gamma^n) 
\int_{x_0- (n+1) L}^{x_0 - n L} a_1(z,\gamma^n) b_0(z)\, \rmd z
+
a_0(\gamma^n) 
\int_{x_0- (n+1) L}^{x_0 - n L} \tilde d_1(z,\gamma^n) b_0(z)\, \rmd z
\nonumber \\
+
a_0(\gamma^n) 
\int_{x_0- (n+1) L}^{x_0 - n L} \tilde d_\Delta(z,\gamma^n) b_1(z,\gamma^n)\, \rmd z
+O(k).
\end{eqnarray}
Note that $F_n(\gamma^n) + k G_n(\gamma^n,k)$ is a part of $D_n(\gamma^n,k)$ of equation (A.27).
Since $\tilde d_\Delta$, $\tilde d_0$ and $\tilde d_1$ include either $V_\Delta(z)$ or $V_\Delta(x_0 - nL)$, both $F_n$ and $G_n$ decrease like $V_\Delta(x_0 - n L)$ as $n \to \infty$. (The $O(k)$ term in (A.76) also decreases like $V_\Delta(x_0 - n L)$.)
So, if $V_\Delta \in F_0^{(-)}$, then, as in (A.28),
\begin{equation}
\fl
\lim_{k \to 0} \sum_{n=0}^\infty [F_n(\gamma^n) + k G_n(\gamma^n,k)]=\sum_{n=0}^\infty \lim_{k \to 0}  [F_n(\gamma^n) + k G_n(\gamma^n,k)] =\sum_{n=0}^\infty F_n(1),
\end{equation}
where we have used $\vert G_n(1,0) \vert < \infty$.
Substituting
\begin{equation}
a_0(1)=\frac{1}{1+c_0},
\qquad
\tilde d_\Delta(z,1)=-\frac{1}{2} (1+c_0)V_\Delta(z)
\end{equation}
and $b_0(z)=g_{\rm p}(z,W)$ in (A.75), we obtain
\begin{equation}
F_n(1)
=
-\frac{1}{2}\int_{x_0- (n+1) L}^{x_0 - n L} V_\Delta(z) g_{\rm p}(z,W)\, \rmd z,
\end{equation}
and hence
\begin{eqnarray}
\fl
\lim_{k \to 0}
\sum_{n=0}^\infty
\int_{x_0- (n+1) L}^{x_0 - n L} a(z,k) \,b(z,k)\, \tilde d(z,k)\, \rmd z
=
\sum_{n=0}^\infty F_n(1)
=
-\frac{1}{2}\int_{-\infty}^{x_0} V_\Delta(z) g_{\rm p}(z,W)\, \rmd z.
\nonumber \\
\end{eqnarray}

Thus we have obtained, for the first term on the right-hand side of (A.37),
\begin{equation}
\lim_{k \to 0}
\int_{-\infty}^{x_0} \frac{-\zeta_\Delta (z)}{\zeta_{\rm p}^2(z)\eta_{\rm p}(z)}
g_{\rm p}(z,\bar \omega_{\rm p}) \, \rmd z
=
-\frac{1}{2} \int_{-\infty}^{x_0} V_\Delta(z) g_{\rm p}(z,W)\, \rmd z.
\end{equation}
Since $g_{\rm p}'(x,W)$ also satisfies (4.10), equation (A.81) holds with $g_{\rm p}$ replaced by $g'_{\rm p}$:
\begin{equation}
\lim_{k \to 0}
\int_{-\infty}^{x_0} \frac{-\zeta_\Delta (z)}{\zeta_{\rm p}^2(z)\eta_{\rm p}(z)}
g'_{\rm p}(z,\bar \omega_{\rm p}) \, \rmd z
=
-\frac{1}{2} \int_{-\infty}^{x_0} V_\Delta(z) g'_{\rm p}(z,W)\, \rmd z.
\end{equation}
In the same way, we can derive
\begin{equation}
\lim_{k \to 0}
\int_{-\infty}^{x_0} \frac{-\eta_\Delta (z)}{\zeta_{\rm p}(z)\eta_{\rm p}^2(z)}
g_{\rm p}(z,\bar \omega_{\rm p}) \, \rmd z
=
\frac{1}{2} \int_{-\infty}^{x_0} V_\Delta(z) g_{\rm p}(z,W)\, \rmd z,
\end{equation}
\begin{equation}
\lim_{k \to 0}
\int_{-\infty}^{x_0} \frac{-\eta_\Delta (z)}{\zeta_{\rm p}(z)\eta_{\rm p}^2(z)}
g'_{\rm p}(z,\bar \omega_{\rm p}) \, \rmd z
=
\frac{1}{2} \int_{-\infty}^{x_0} V_\Delta(z) g'_{\rm p}(z,W)\, \rmd z.
\end{equation}
It is easy to see that
\begin{eqnarray}
\fl
\lim_{k \to 0}
\int_{-\infty}^{x_0} \frac{V_\Delta (z)}{\zeta_{\rm p}(z)\eta_{\rm p}(z)}
g'_{\rm p}(z,\bar \omega_{\rm p}) \, \rmd z
=
\int_{-\infty}^{x_0}
 \lim_{k \to 0}
\frac{V_\Delta (z)}{\zeta_{\rm p}(z)\eta_{\rm p}(z)}
g'_{\rm p}(z,\bar \omega_{\rm p}) \, \rmd z
\nonumber \\
=
\int_{-\infty}^{x_0} V_\Delta(z) g'_{\rm p}(z,W)\, \rmd z.
\end{eqnarray}
Taking the limit $k \to 0$ of (A.37), and substituting (A.81)--(A.85), we obtain (A.13). 
This conclusion does not change when the terms of higher order in $V_\Delta$ are taken into account.

\section{Proof of (7.15)}
Since $U(z,x;k)$ is the inverse of $U(x,z;k)$, for $z \leq x$ we have $\alpha(z,x;\pm k)=\alpha(x,z;\mp k)$ and $\beta(z,x;\pm k)=-\beta(x,z;\pm k)$.
Using this, from (2.1) we obtain
\begin{eqnarray}
\frac{\partial}{\partial z}\bar \alpha(x,z;W;k)
=\rmi k \bar \alpha(x,z;W;k) - f(z) \bar \beta(x,z;W;-k),
\nonumber \\
\frac{\partial}{\partial z}\bar \beta(x,z;W;-k)
=-\rmi k \bar \beta(x,z;W;-k) - f(z) \bar \alpha(x,z;W;k),
\end{eqnarray}
where $\bar \alpha$ and $\bar \beta$ are defined by (A.2) of appendix~A. 
Hence, for ${\rm Im}\,k \geq 0$,
\begin{eqnarray}
\fl
\frac{\partial}{\partial z}
\left(
\vert \bar \alpha(x,z;W;k)\vert^2-\vert \bar \beta(x,z;W;-k) \vert^2
\right)
\nonumber \\
=-2 ({\rm Im}\,k)
\left(
\vert \bar \alpha(x,z;W;k)\vert^2+\vert \bar \beta(x,z;W;-k) \vert^2
\right) \leq 0.
\end{eqnarray}
Since $\vert \bar \alpha(x,x;W;k)\vert^2-\vert \bar \beta(x,x;W;-k) \vert^2=1$, from (B.2) it follows that 
\begin{equation}
\vert \bar \alpha(x,z;W;k)\vert^2-\vert \bar \beta(x,z;W;-k) \vert^2 \geq 1
\end{equation}
for $z \leq x$. Using (A.3) of appendix~A, we have
\begin{equation}
\vert \bar \tau(x,z;W;k) \vert^2 + \vert \bar R_l(x,z;W;k) \vert^2 \leq 1,
\end{equation}
and this gives
\begin{equation}
\vert \bar Q(x,z;W;k) \vert =
\left\vert
\frac{\bar \tau^2(x,z;W;k)}{1-\bar R_l^2(x,z;W;k)}
\right\vert
\leq
\frac{\vert \bar \tau(x,z;W;k) \vert^2}{1-\vert \bar R_l(x,z;W;k) \vert^2}
\leq 1.
\end{equation}

\section{Exact Green function for $\boldsymbol{V_{\rm S}}$ given by (12.23)}
Let us define
\begin{eqnarray}
\fl
p_k \equiv \sqrt{k^2+E_0}, \qquad
q_k \equiv \sqrt{C- k^2 - E_0}, \qquad
s_k \equiv \sqrt{h- k^2 - E_0},
\nonumber \\
\fl
\alpha \equiv \rme^{q_k b}
\left(\cos p_k a + \frac{q_k^2-p_k^2}{2 p_k q_k} \sin p_k a\right), \qquad
\alpha' \equiv \rme^{-q_k b}
\left(\cos p_k a - \frac{q_k^2-p_k^2}{2 p_k q_k} \sin p_k a\right),
\nonumber \\
\fl
C_1 \equiv \frac{-C}{2 p_k q_k}\rme^{-q_k b} \sin p_k a, \qquad
C_2 \equiv \frac{-\alpha + \alpha' + \rmi \sqrt{4-(\alpha+\alpha')^2}}{2}.
\end{eqnarray}
The exact $G_{\rm S}(x,y;k)$ for (12.23) ($0<y \leq x<a$) can be obtained as
\begin{equation}
G_{\rm S}(x,y;k)=-\frac{\psi^+(x) \psi^-(y)}{W[\psi^+, \psi^-]},
\end{equation}
where
\begin{eqnarray}
\fl
\psi^+(x)\equiv (C_1+C_2) \cosh [s_k (x-a)] + (q_k/s_k)(C_1-C_2)\sinh [s_k(x-a)], 
\nonumber \\
\fl
\psi^-(x)\equiv (C_1+C_2) \cosh s_k x - (q_k/s_k)(C_1-C_2)\sinh s_k x, 
\end{eqnarray}
\begin{eqnarray}
\fl
W[\psi^+,\psi^-]=\left[(C_1+C_2)^2
+ (q_k/s_k)^2 (C_1-C_2)^2\right]s_k \sin s_k a
-2 q_k(C_1^2 - C_2^2) \cos s_k a.
\nonumber \\*
\end{eqnarray}


\end{document}